\documentclass[prd,twocolumn,superscriptaddress,
showpacs,showkeys,preprintnumbers] {revtex4}

\usepackage{graphicx} \usepackage{float} \usepackage{color}
\usepackage{natbib} \bibpunct{(}{)}{;}{a}{,}{,}
\usepackage{setspace}
\usepackage{natbib}
\bibpunct{[}{]}{,}{a}{,}{,}

\usepackage{amsmath,amssymb, amsfonts}

\usepackage{subfigure}

\usepackage{multirow}

 \newcommand{\ket}[1]{
  |{#1}\rangle}

\begin{document}

\title{Phenomenological analysis of quantum collapse as
  source of the seeds of cosmic structure}
  \author{Adolfo \surname{De Un\'{a}nue}}
  \email[E-mail me at: ]{adolfo@nucleares.unam.mx}
  \affiliation{Instituto de Ciencias
    Nucleares, Universidad Nacional Aut\'{o}noma de
    M\'{e}xico, \\ 04510 Ciudad de M\'{e}xico, M\'{e}xico.}
  \author{Daniel \surname{Sudarsky}}
  \email[E-mail me at:]{sudarsky@nucleares.unam.mx}
  \affiliation{Instituto de Ciencias
    Nucleares, Universidad Nacional Aut\'{o}noma de
    M\'{e}xico, \\ 04510 Ciudad de M\'{e}xico, M\'{e}xico.}

\begin{abstract}
  The standard inflationary version of the origin of the
  cosmic structure as the result of the quantum fluctuations
  during the early universe is less than fully satisfactory
  as has been argued in [A.~Perez, H.~Sahlmann, and
  D.~Sudarsky, Class. Quantum Grav., \textbf{23}, 2317,
  (2006)]. A proposal is made there of a way to address the
  shortcomings by invoking a process similar to the collapse
  of the quantum mechanical wave function of the various
  modes of the inflaton field. This in turn was inspired on
  the ideas of R. Penrose about the role that quantum
  gravity might play in bringing about such breakdown of the
  standard unitary evolution of quantum mechanics. In this
  paper we study in some detail the two schemes of collapse
  considered in the original work together with an
  alternative scheme, which can be considered as ``more
  natural'' than the former two. The new scheme, assumes
  that the collapse follows the correlations indicated in
  the Wigner functional of the initial state.  We end with
  considerations regarding the degree to which the various
  schemes can be expected to produce a spectrum that
  resembles the observed one.
\end{abstract}

\date{January 30, 2008}

\maketitle

\section{Introduction}

In recent years, there have been spectacular advances in
physical cosmology, resulting from remarkable increase in
the accuracy of the observational techniques and exemplified
by the Supernova Surveys \cite{Supernovas}, the studies of
large scale structure \cite{Structure} and the highly
accurate observations from various recent studies in
particular those of Wilkinson Microwave Anisotropy Probe
(WMAP) \cite{wmap2007}. These observations have strengthened
the theoretical status of the Inflationary scenarios among
cosmologists.

We should note however that while much of the focus of the
research in Inflation has been directed towards the
elucidation of the exact form of the inflationary model
(i.e. the number of fields, the form of the potential, and
the occurrence of non-minimal couplings to gravity to name a
few), much less attention has been given to the questions of
principle, as how the initial conditions are determined,
what accounts for the low entropy of the initial state, and
how exactly does the universe transit from a homogeneous and
isotropic stage to one where the quantum uncertainties
become actual inhomogeneous fluctuations. There are of
course several works in which this issues are addressed
\cite{ Other, Others} but as explained in \cite{Perez2006,
  Sudarsky06b, Sudarsky07} the fully satisfactory account of
the last of them seems to require something beyond the
current understanding of the laws of physics.  The point is
that the predictions of inflation in this regard can not be
fully justified in any known and satisfactory
interpretational scheme for quantum physics.  The Copenhagen
interpretation, for instance, is inapplicable in that case,
due to the fact that we, the observers, are part of the
system, and to make things even worse we are in fact part of
the outcome of the process we wish to understand, Galaxies,
Stars, planets and living creatures being impossible in a
homogeneous and isotropic universe \footnote{Further issues
  about the identification of ensemble averages with time
  averages, which are valid under the ergodic hypothesis can
  be raised in the cosmological context where assumptions of
  ergodicity and equilibrium seem to be much less
  justified.}.
The arguments and counter-arguments that have arisen in
regard to this aspect of the article mentioned above have
been discussed in various other places by now and we point
the reader who is interested in that debate to that
literature \cite{debate, Sudarsky07}.  In the present work
we will focus on a more detailed study of the collapse
schemes and on the traces they might leave on the
observational data.  Nevertheless, and in order to make the
article self contained, we will briefly review the
motivation and line of approach described in detail in
\cite{Perez2006}.

To clarify where lies the problem, and the way in which it
is addressed in \cite{Perez2006} we will review in a
nutshell, the standard explanation of the origin of the
seeds of cosmic structure in the Inflationary paradigm:
\begin{itemize}
\item One starts with an homogeneous and isotropic
  spacetime\footnote{Inflation could work if we don't start
    -strictly- with this condition, but after some
    e-foldings the universe reaches this stage.} The
  inflaton field is the dominant matter in this spacetime,
  and it is in its vacuum quantum state,
  which is \emph{homogeneous and isotropic too}.  The field
  is in fact described in terms of its expectation value
  represented as a scalar field which depends only on cosmic
  time but not on the spatial coordinates, $\phi_0$ and a
  quantum or ``fluctuating part'' $ \delta\phi$ which is in
  the adiabatic vacuum state, which is an homogeneous and
  isotropic state (something that can be easily verified by
  applying the generators of rotations or translations to
  the state).
\item The quantum ``fluctuations'' of the inflaton acts as
  perturbations \footnote{ We find this wording unfortunate
    because it leads people to think that something is
    fluctuating in the sense of Brownian motion, while a
    wording such as "quantum uncertainties'' would evoke
    something like the wave packet associated with a ground
    state of an harmonic oscillator which is a closer
    analogy with what we have at hand.} of the inflaton
  field and through the Einstein Field Equations (EFE) as
  perturbations of the metric.
\item As inflation continues the physical wave length of the
  various modes of the inflaton field become larger than the
  Hubble Radius (Horizon-crossing as referred commonly in
  the literature), and the quantum amplitudes of the modes
  freeze.  At that moment one starts regarding such modes as
  actual waves of in a classical field.  Later on, after
  inflation ends, and as the Hubble Radius grows, the
  fluctuations ``re-enter the Horizon'', transforming at
  that point into the seeds of the cosmic structure.
\end{itemize}
The last step is usually refereed as the quantum to
classical transition. There are of course several schools of
though about the way one must consider such transition: from
those using the established physical paradigms \cite{Other,
  Others}, to views advocating a certain generalization of
the standard formalisms\cite{Hartle}.  The two works
\cite{Other, Hartle} focus concretely in a full blown
quantum cosmology, and its interpretational problems, which
are even more severe than the ones we are dealing with
here. In \cite{Perez2006} it was argued that such schemes
are insufficient, in particular if one expects cosmology to
provide a time evolution account starting from the totally
symmetric state to an inhomogeneous and anisotropic Universe
in which creatures such as humans might eventually arise.

The view taken in \cite{Perez2006} (and in this work)
intends to be faithful to the notion that physics is always
quantum mechanical, and that the only role for a classical
description is that of an approximation where the
uncertainties in the state of the system are negligibly
small and one can take the expectation values as fair
description of the aspects of the state one is interested
on. However one must keep in mind that behind any classical
approximation there should always lie a full quantum
description, and thus one should reject any scheme in which
the classical description of the universe is inhomogeneous
and anisotropic but in which the quantum mechanical
description persists in associating to the universe an
homogeneous and isotropic state at all times.
Thus in \cite{Perez2006}, one introduces a new ingredient to
the inflationary account of the origin of the seeds of
cosmic structure: \emph{the self induced collapse
  hypothesis}. I.e. one considers a specific scheme by which
a self induced collapse of the wave function is taken as the
mechanism by which inhomogeneities and anisotropies arise in
each particular scale. This work was inspired in early ideas
by Penrose \cite{Penrose} which regard that the collapse of
the wave function as an actual physical process (instead of
just an artifact of our description of physics) and which is
assumed to be caused somehow by quantum aspects of
gravitation. We will not recapitulate the motivations and
discussion of the original proposal and instead refer to the
reader to the above mentioned works.

The way we treat the transition of our system from a state
that is homogeneous and isotropic to one that is not, is to
assume that at certain cosmic time, something induces a jump
in a state describing a particular mode of the quantum
field, in a manner that would be similar to the standard
quantum mechanical collapse of the wave function associated
with a measurement, but with the difference that in our
scheme no external measuring device or observer is called
upon as ``triggering'' that jump (it is worthwhile recalling
that nothing of that sort exist, in the situation at hand,
to play such role).

The main aim of this article is to compare the results that
emerge from the collapse schemes considered in
\cite{Perez2006} with an alternative scheme of collapse that
can be said to more natural than the previous two.  In this
new scheme \footnote{As we will show, the relevant
  quantities that one is interested in computing are
  determined once one characterizes the time of collapse and
  describes the state after the collapse in terms of the
  expectation values of the field and momentum conjugate
  variables.} we take into account the correlations in the
quantum state of the system before the collapse for the
values of field and conjugate momentum variables as
indicated by the Wigner functional analysis of the
pre-collapse state.

This article is organized as follows: In the Section
\ref{sec:preliminares} we review the formalism used in
analysing the collapse process. Section
\ref{sec:collapse_wigner} review how to obtain the wave
function for the field from its Fock space description,
which is then used in evaluating the Wigner function for the
state, and the state that results after the collapse.
Section \ref{sec:comp_observations} describes the details of
the spectrum of cosmic fluctuations, resulting from such
collapse, and finally, in Section \ref{sec:discussion} we
discuss these results and those of other collapse schemes
\emph{vis a vie} the empirical data.

\section{\label{sec:preliminares}The Formalism}
The starting point is the action of a scalar field with
minimal coupling to the gravity sector:
\begin{multline}
  \label{eq:action}
  S[\phi, g_{ab}] = \int d^4x\sqrt{-g}\,\Big( \frac{1}{16\pi
    G} R[g_{ab}]- \\ \frac{1}{2}\nabla_a\phi\nabla_b\phi
  g^{ab} -V(\phi)\Big).
\end{multline}
One splits the corresponding fields into their homogeneous
(``background'') part and the perturbations
(``fluctuation''), so the metric and the scalar field are
written as : $g = g_0 + \delta g$ and $\phi = \phi_0 +
\delta\phi$. With the appropriate choice of gauge (conformal
Newton gauge) and ignoring the vector and tensor part of the
metric perturbations, the space-time metric is described by.
\begin{equation}
  \label{eq:conformal-newton}
  ds^2 = a(\eta)^2\left[ -(1+2\Psi)d\eta^2 + (1-2\Psi)
    \delta_{ij}\, dx^i dx^j \right],
\end{equation}
where $\Psi$ is called the \emph{Newtonian potential}.  One
then considers the EFEs to zeroth and first order. The
zeroth order gives rise the standard solutions in the
inflationary stage, where $a(\eta) = -\frac{1}{H_I\eta}$,
with $H_I^2 \simeq (8\pi/3)GV$ with the scalar potential,
$\phi_0$ in slow-regime so $\phi_0' \simeq
-\frac{1}{3H_I}\frac{dV}{d\phi}$; and the first order EFE
reduce to an equation relating the gravitational
perturbation and the perturbation of the field
\begin{equation}
  \label{eq:classical_fundamental}
  \nabla^2\Psi = 4\pi G\phi_0'\delta\phi' \equiv s\delta\phi',
\end{equation}
with $s \equiv 4\pi G\phi_0' $.  The next step involves
quantazing the fluctuating part of inflaton field. In fact
it is convenient to work with the rescaled field $y =
a\delta\phi$. In order to avoid infrared problems we
consider restriction of the system to a box of side $L$,
where we impose, as usual, periodic boundary conditions. We
thus write the fields as
\begin{equation}
  \hat{y}(\eta, \vec{x}) = \frac{1}{L^3}\sum_k e^{i\vec k \cdot\vec x}
  \hat y_k(\eta), \quad   \hat{\pi}(\eta, \vec{x}) =  \frac{1}{L^3}
  \sum_k e^{i\vec k \cdot\vec x}\hat \pi_k(\eta),
\end{equation}
where $\hat\pi_k$ is the canonical momentum of the scaled
field. The wave vectors satisfy $k_iL = 2\pi n_i$ with $i =
1,2,3$, and $\hat y_k(\eta) \equiv y_k(\eta)\hat a_k + \bar
y_k(\eta) \hat a_k^\dag$, and $\hat \pi_k(\eta) \equiv
g_k(\eta)\hat a_k + \bar\pi_k(\eta)\hat a_k^\dag$. The
functions $y_k(\eta), g_k(\eta)$ reflect our election of the
vacuum state, the so called Bunch-Davies vacuum:
\begin{equation}
  \label{eq:Bunch-Davies}
  y_k(\eta) = \frac{1}{\sqrt{2k}} \left( 1 - \frac{i}{\eta
      k }\right) e^{ -i k \eta }, \quad
  g_k(\eta) = -i  \sqrt {\frac{k}{2}} e^{- i k \eta }.
\end{equation}

The vacuum state is defined by the condition $\hat a_k
|0\rangle =0$ for all $k$, and is homogeneous and isotropic
at all scales.  As indicated before, according to the
proposal, the self-induced collapse operates in close
analogy with a ``measurement'' in the quantum-mechanical
sense, and assumes that at a certain time $\eta^c_k$ the
part of the state that describes the mode $\vec k$ jumps to
an new state, which is no longer homogeneous and isotropic.
To proceed to the detailed description of this process, one
decomposes the fields into their hermitian parts as follows
$\hat y_k = \hat y_k^R(\eta) + i \hat y_k^I(\eta)$, and
$\hat \pi_k = \hat \pi_k^R(\eta) + i \hat \pi_k^I(\eta)$.

We note that the vacuum state $\ket{0}$ is characterized in
part by the following: its expectation values $ \langle \hat
y_k^{R,I}(\eta)\rangle = \langle \hat \pi_k^{R,I}(\eta)
\rangle=0$ and its uncertainties are $\Delta\hat y_k^ {R,I}
= 1/2|y_k|^2(\hbar L^3)$ and $\Delta\hat \pi_k^{R,I} =
1/2|g_k|^2(\hbar L^3)$.

For an arbitrarily given state of the field $\ket{\Omega}$,
we introduce the quantity $d_k \equiv \langle \Omega | \hat
a_k^{R,I}|\Omega \rangle \equiv |d_k^{R,I}| e^{i \alpha_k}$
so that,
\begin{equation}
  \label{eq:expectation_values}
  \langle \hat y_k^{R,I}\rangle = \sqrt 2 \, \Re \, (y_k
  d_k^{R,I}), \quad \langle \hat \pi_k^{R,I} \rangle = \sqrt 2 \,
  \Re \, (g_k d_k^{R,I} ),
\end{equation}
which shows that it specifies the main quantity of interest
in characterizing the state of the field.

It is convenient for future use to define the following
phases, $\beta_k = \arg(y_k)$ and $\gamma_k = \arg(g_k)$,
keeping in mind that they depend on the conformal time $\eta
$.

The analysis now calls for the specification of the scheme
of collapse determining the state of the field after the
collapse\footnote{ At this point, in fact, all we require is
  the specification of the expectation values of certain
  operators in this new quantum state. }, which is the main
purpose of the next section.  With such collapse scheme at
hand one then proceeds to evaluate the perturbed metric
using a semi-classical description of gravitation in
interaction with quantum fields as reflected in the
semi-classical EFE's: $G_{ab} = 8\pi G \langle T_{ab}
\rangle$.  To lowest order this set of equations
reduces to
\begin{equation}
  \label{eq:semiclassical-fundamental}
  \nabla^2\Psi_k = s\langle \delta\hat\phi'_k \rangle_\Omega,
\end{equation}
where $\langle \delta\hat\phi'_k \rangle_\Omega$ is the
expectation value of the momentum field $\delta\hat\phi_k' =
\hat\pi_k/a(\eta)$ on the state $\ket{\Omega}$
characterizing the quantum part of the inflaton field.  It
is worthwhile emphasizing that \emph{before} the collapse
has occurred \emph{there are not} metric
perturbations\footnote{This might seem awkward to some
  readers. It is worth then emphasizing that our view is
  that, in contrast with what happens with other fields, the
  fundamental degrees of freedom of gravitation are not
  related to the metric degrees of freedom in any simple
  way, but instead the latter appear as effective degrees of
  freedom of a non-quantum effective theory.  Therefore, the
  quantum uncertainties (we feel "uncertainties" is a more
  appropriate word than "fluctuations", as the latter
  suggest that something is actually changing constantly in
  a random way) associated with the gravitational degrees of
  freedom are most naturally thought as not having a metric
  description (as occurs for instance in the Loop Quantum
  Gravity program where the fundamental degrees of freedom
  are holonomies and fluxes), and thus that the metric can
  appear only at the classical level of description, where
  it satisfies something close to the semiclassical Einstein
  equation.  In other words, from our point of view, it
  would be incorrect to think of the quantum uncertainties
  of the metric as appropriate description of the quantum
  aspects of gravitation, and much less, as satisfying
  Einstein's equations.  From our point of view, this would
  be analogous to imagining the quantum indeterminacies
  associated with the ground state of the hydrogen atom, as
  described in terms a perturbation of the orbit of an
  electron in Hydrogen atom, and satisfying Keppler's
  equations for the classical Coulomb potential.  For more
  details about these point of view see
  \cite{Perez2006,Sudarsky06b,Sudarsky07}.  The reader
  should be aware that this is not a view shared by most
  cosmologists.}, i.e. the r.h.s. of the last equation is
zero, so, it is only \emph{after} the collapse that the
gravitational perturbations appear, i.e. the collapse of
each mode represents the onset of the inhomogeneity and
anisotropy at the scale represented by the mode.  Another
point we must stress is that, after the collapse,and in fact
at all times, our \emph{Universe would be defined by a
  single state $\ket\Omega$, and not by an ensemble of
  states}.  The statistical aspects arise once we note that
we do not measure directly and separately each the modes
with specific values of $\vec k$, but rather the aggregate
contribution of all such modes to the spherical harmonic
decomposition of the temperature fluctuations of the
celestial sphere (see below).

To make contact with the observations we note that the
quantity that is experimentally measured (for instance by
WMAP) is $\Delta T/ T (\theta, \varphi)$, which is expressed
in terms of its spherical harmonic decomposition
$\sum_{lm}\alpha_{lm}Y_{lm}(\theta,\varphi)$.  The contact
with the theoretical calculations is made trough the
theoretical estimation most likely value of the
$\alpha_{lm}$'s, which are expressed in terms of the
Newtonian potential on the 2-sphere corresponding to the
intersection of our past light cone with the of last
scattering surface (LSS): $\Psi(\eta_D, \vec x_D)$, $\alpha_{lm} =
\int \Psi(\eta_d, \vec x_D) Y_{lm}^* d^2\Omega$. We must
then consider the expression for the Newtonian Potential
(\ref{eq:semiclassical-fundamental}) at those points:
\begin{equation}
  \label{eq:fundamental}
  \Psi(\eta, \vec x) = \sum_k \frac{s \mathcal{T}(k)}{k^2L^3} \langle
  \delta\hat\phi'_k \rangle
  e^{i\vec k \cdot \vec x},
\end{equation}
where we have introduced the factor $\mathcal{T}(k)$ to represent the
physics effects  of the period between reheating and
decoupling.

Writing the coordinates of the points of interest on the
surface of last scattering as $\vec x = R_D(\sin\theta
\sin\phi, \sin\theta \cos\phi, \cos\theta)$, where $R_D$ is
the comoving radius of that surface and $\theta, \phi$ are
the standard spherical coordinates of the sphere, and using
standard results connecting Fourier and spherical expansions
we obtain
\begin{align}
  \alpha_{lm} &= \sum_k\frac{s \mathcal{T}(k)}{k^2L^3} \int
  \langle\delta\hat\phi_k'\rangle e^{i\vec k \cdot \vec x}
  Y_{lm}(\theta,\phi) d^2\Omega \\
  &= \frac{s}{L^3}\sum_k\frac{\mathcal{T}(k)}{k^2}
  \langle\delta\hat\phi_k'\rangle 4\pi i^lj_l(|\vec
  k|R_D)Y_{lm}(\hat k).
\end{align}
As indicated above statistical considerations arise when
noting that the equation (\ref{eq:fundamental}) indicates
that the quantity of interest is in fact the result of a
large number (actually infinite) of harmonic oscillator each
one contributing with a complex number to the sum, leading
to what is in effect a two dimensional random walk whose
total displacement corresponds to the observational
quantity.  Note that this part of the analysis is
substantially different from the corresponding one in the
standard approach.  In order to obtain a prediction, we need
to find the most likely value of the {\it magnitude} of such
total displacement.

Thus we must concern ourselves with:
\begin{multline}\label{alpha_cuadrado}
  |\alpha_{lm}|^2 = \frac{16 s^2
    \pi^2}{L^6}\sum_{\vec{k}\vec{k}'} \frac{\mathcal{T}(k)}{k^2}
  \frac{\mathcal{T}(k')}{k'^2} \times \\
  \langle\delta\hat\phi_k'\rangle
  \langle\delta\hat\phi_{k'}'\rangle^* j_l(kR_D)j_l(k'
  R_D)Y_{lm}(\hat{k})Y_{lm}(\hat{k}'),
\end{multline}
and to obtain the ``most likely'' value for this quantity.
This we do with the help of the {\it imaginary} ensemble of
universes\footnote{This is just a mathematical evaluation
  device and no assumption regarding the existence of such
  ensemble of universes is made or needed. These aspects of
  our discussion can be regarded as related to the so called
cosmic variance problem.} and the
identification of the most likely value with the ensemble
mean value.

As we will see, the ensemble mean value of the product
$\langle\delta\hat\phi_k\rangle
\langle\delta\hat\phi_{k'}\rangle^*$, evaluated in the
post-collapse states \footnote{ Note here again the
  difference with the standard treatment of this part of the
  calculation which calls for the evaluation of the
  expectation value
  $\langle\delta\hat\phi_k\delta\hat\phi_{k'}\rangle^*$ on
  the vacuum state which as already emphasized is completely
  homogeneous and isotropic.}, results in a form $\kappa C(k)
\delta_{\vec{k}\vec{k}'} $, where $\kappa = \hbar L^3 k / (4
a^2)$ and $C(k)$ is an adimensional function of $k$ which
codifies the traces of detailed aspects of the collapse
scheme.\ We are thus lead to the following expression for
the most likely (ML) value of the quantity of interest:
\begin{equation}
  |\alpha_{lm}|^2_{ML} = s^2\frac{4 \pi^2 \hbar}{L^3 a^2}
  \sum_{\vec{k}}
  \frac{ C(k)  \mathcal{T}(k)^2}{k^3} j_l^2(|\vec{k}|R_D)|Y_{lm}(\hat{k})|^2.
\end{equation}
Writing the sum as an integral (using the fact that the
allowed values of the components of $\vec k$ are separated
by $\Delta k_i = 2\pi /L$):
\begin{equation}
  |\alpha_{lm}|^2_{ML}  = \frac{s^2 \hbar }{2 \pi a^2}
  \int   \frac{C(k) \mathcal{T}(k)^2}{k^3}
  j_l^2(|\vec{k}|R_D)|Y_{lm}(\hat{k})|^2 d^3k.
\end{equation}

The last expression can be made more useful by changing the
variables of integration to $x = k R_D$, leading to
\begin{equation}\label{eq:contacto_obs}
  |\alpha_{lm}|^2_{ML}  = \frac{s^2\hbar}{2 \pi a^2}
  \int   \frac{C(x/R_D)}{x} \mathcal{T}(x/R_D)^2
  j_l^2(x)  dx.
\end{equation}
With this expression at hand we can compare the expectations
from each of the schemes of collapse against the
observations. We note, in considering the last equation,
that the standard form of the spectrum corresponds to
replacing the function $C$ by a constant. In fact if one
replaces $C$ by $1$ and one further takes the function $\mathcal{T}$
which encodes the late time physics including the plasma
oscillations which are responsible for the famous acoustic
peaks, and substitutes it by a constant, one obtains the
characteristic signature of a scale invariant spectrum: $
|\alpha_{lm}|^2_{ML} \varpropto \frac{1}{l(l+1)}$.

In the remaining of the paper we will focus on the effects
that a nontrivial form of the function $C$ has on the
predicted form of the observational spectrum.

\section{\label{sec:collapse_wigner}Proposal of Collapse
  \lowercase{\emph{a l\`a}} Wigner}

As indicated in the introduction, the schemes of collapse
considered in the first work following the present approach,
\cite{Perez2006}, essentially ignored the correlations
between the canonical variables that are present in the
pre-collapse vacuum state.  In the present analysis, we will
focus on this feature, characterising such correlations via
the Wigner distribution function \cite{Wigner1932}, and
requiring the collapse state to reflect those aspects.  The
choice of the Wigner distribution function to describe these
correlations in this setting is justified by some of its
standard properties regarding the "classical limit'' (see
for instance \cite{Ballentine2000}), and, by the fact that
there is a precise sense in which it is known to encodes the
correlations in question \cite{Wigner}. The Wigner
distribution function for pure quantum states characterized
by a position space wave function $\Psi(q)$ is defined as:
\begin{equation}
  \label{eq:wigner_function}
  \mathcal{W}(q,p) =
  \frac{1}{2\pi\hbar}\int_{-\infty}^{\infty} d\,y
  \Psi^*(q+y)\Psi(q-y) \exp{\left({\frac{ipy}{\hbar}}\right)},
\end{equation}
with $(q,p)$ corresponding to the canonical conjugate
variables.

In our case the wave function for each mode of the field
(characterized by its wave vector number $\vec k$)
corresponds, initially, to the ground state of an harmonic
oscillator.  It is a well known result that the Wigner
distribution function gives for a quantum harmonic
oscillator in its vacuum state a bi dimensional Gaussian
function. This fact will be used to model of the result of
collapse of the quantum field state.  The assumption will be
that at a certain (conformal) time $\eta_k^c$ the part of
the state characterizing the mode $k$, will collapse (in a
way that is similar to what in the Copenhagen interpretation
is associated with a measurement), leading to a new state
$\ket{\Omega}$ in which the fields (expressed by its
hermitian parts) will have expectation values given by
\begin{equation}
  \label{eq:valores_medios_wigner}
  \langle\hat{y}_k^{R,I}\rangle_\Omega  = x^{(R,I)}_k \Lambda_k
  \cos\Theta_k\, ,
  \quad
  \langle\hat{\pi}_k^{R,I}\rangle_\Omega  = x^{(R,I)}_k \Lambda_k k
  \sin\Theta_k,
\end{equation}
where $x^{(R,I)}$ is a random variable, characterized by a
Gaussian distribution centered at zero with a spread one;
$\Lambda_k$ is given by the major semi-axis of the ellipse
characterizing the bi dimensional Gaussian function (the
ellipse corresponds to the boundary of the region in ``phase
space'' where the Wigner function has a magnitude larger
than $ 1/2$ its maximum value), and $\Theta_k $ is the angle
between that axis and the $y_k^{R,I }$ axis.

Comparing \eqref{eq:expectation_values} with
(\ref{eq:valores_medios_wigner}) we obtain,
\begin{align}
  |d_k^{R,I}|\cos(\alpha_k+\beta_k) &=
  \frac{1}{\sqrt{2}|y_k|}x_k^{R,I} \Lambda_k
  \cos\Theta_k, \\
  |d_k^{R,I}|\cos(\alpha_k+\gamma_k) &=
  \frac{1}{\sqrt{2}|g_k|}x_k^{R,I} \Lambda_k k \sin\Theta_k.
\end{align}
From this expressions we can solve for the constants
$d_k^{R,I} = |d_k^{R,I}| e^{i \alpha_k}$.  In fact using the
polar representation of the $y_k$ and $g_k$ we find
\begin{equation}
  \tan(\alpha_k -k\eta) = \frac{k^2\eta^c |y_k|
    \sin\Theta_k}{|g_k|\cos\Theta_k\sqrt{1+k^2(\eta^c)^2} - k|y_k|\sin\Theta_k}
\end{equation}
obtaining
\begin{multline}
  \label{eq:d_k}
  |d_k^{R,I}| = \frac{x_k^{R,I} \Lambda_k}{\sqrt{2}
    |y_k||g_k|}\cdot\frac{\sqrt{1+k^2\eta_c^2}}{k\eta_c} \times \\
  {\sqrt{|y_k|^2 k^2\sin^2\Theta_k+|g_k|^2\cos^2\Theta_k-
      \frac{2|y_k||g_k| k
        \cos\Theta_k\sin\Theta_k}{(1+k^2\eta_c^2)^{1/2}}}},
\end{multline}
where in all of the expressions above the conformal time
$\eta$ is set to the time of collapse $\eta_c^k$ of the
corresponding mode.

In order to obtain the expression for $\Lambda_k$ it is
necessary to find the wave-function representation of the
vacuum state for the variable $y_k^{R,I }$.  Following a
standard procedure, we apply the annihilation operator,
$\hat{a}^{R,I}$, to the vacuum state $\ket{0}$, obtaining
the well-known equation for the harmonic oscillator in the
vacuum state, and from the result we extract the wave
function of the $k-$mode of the inflaton field:
\begin{multline}
  \label{eq:wave_function}
  \Psi^{R,I}\left(y_k^{R,I}, \eta\right) =
  \left(\frac{2k}{\left(1+\displaystyle
        \frac{i}{k\eta}\right)\pi\hbar L^3}\right)^{1/4}
  \times \\
  \exp\left({-\frac{k}{\hbar L^3\left(1 + \displaystyle
          \frac{i}{k\eta}\right)}\left(y_k^{R,I}\right)\,^2}\right).
\end{multline}
We next substitute this in the expression for the Wigner
function, $\mathcal{W}(y_k^{R,I}, \pi_k^{R,I}, \eta)$,
obtaining,
\begin{multline}
  \mathcal{W}(y_k^{R,I},\pi_k^{R,I},\eta) = \\
  2\left(1+\frac{1}{k^2\eta^2}\right)^{1/4}
  \exp\left(-\frac{2k}{\hbar
      L^3}\left(y_k^{R,I}\right)^2\right) \times \\
  \exp\left(\frac{2}{k\eta\hbar
      L^3}y_k^{R,I}\pi_k^{R,I}\right)
  \exp\left(-\frac{(1+k^2\eta^2)}{2\hbar L^3
      k^3\eta^2}\left(\pi_k^{R,I}\right)^2\right).
\end{multline}
This has the form of a bi dimensional Gaussian distribution
as expected from the form of the vacuum state. The cross
term is telling us that the support of Wigner function is
rotated respect the original axes.  Rescaling the
$\pi_k$-axe to $\Pi_k = \pi_k/k$ and doing a simple 2D
rotation (i.e. $y_k'^{R,I} = y_k^{R,I} \cos \Theta_k +
\Pi_k^{R,I}\sin \Theta_k $, $\Pi_k'^{R,I} =
\Pi_k^{R,I}\cos\Theta_k - y_k^{R,I} \sin \Theta_k$) we find
the principal axes of the Wigner function:
\begin{multline}
  \label{eq:wigner_rotada}
  \mathcal{W}'(y'\,_k^{R,I}, \Pi'\,_k^{R,I}, \eta) =
  2\left(1+\frac{1}{k^2\eta^2}\right)^{1/4} \times \\ \exp
  \left(-\left(\frac{{y'}_k^{R,I}}{\sigma_{y_k}'}\right)^2\right)
  \exp\left(-\left(\frac{{\Pi'}_k^{R,I}}{\sigma_{\Pi_k'}}\right)^2\right),
\end{multline}
with the corresponding widths given by:
\begin{equation}
  \sigma_{y_k'} = \frac {4 \hbar L^3 k \eta^2}{1+5 k^2\eta^2 + \sqrt
    {1+10 k^2\eta^2 + 9k^4\eta^4}},
\end{equation}
\begin{equation}
  \sigma_{\Pi_k'}= \frac {4 \hbar L^3 k \eta^2}{1+5 k^2\eta^2 - \sqrt
    {1+10 k^2\eta^2 + 9k^4\eta^4}}.
\end{equation}
Note that $\sigma_{\Pi_k'} > \sigma_{y_k'} $. The rotation
angle, $\theta_k$ is given by
\begin{equation}
  2  \Theta_k  = \arctan\left(\frac{4k\eta}{1-3k^2\eta^2}\right).
\end{equation}

It is clear then that $\Lambda_k \equiv 2\sigma_{\Pi_k'}$.

Substituting $\hat\pi_k$ in $\delta\hat\phi'_k$ (defined by
the equation \eqref{eq:fundamental}) and calculating the
expectation value of it in the post-collapse state,
$\ket{\Omega}$, we obtain
\begin{multline}
  \langle\delta\hat\phi'_k\rangle_\Omega=\sqrt{\frac{k}{2}}\cdot\frac{1}{a}
  \Big[\ |d_k^R|\cos\left(\alpha_k^R + \gamma_k +
    \Delta_k\right)
  + \\
  i|d_k^I|\cos\left(\alpha_k^I + \gamma_k +
    \Delta_k\right)\Big],
\end{multline}
where we have defined the ``collapse to observation delay''
from the collapse time of the mode $k$, $\eta^c_k$ as
$\Delta_k = k(\eta -\eta^c_k)$ where $\eta$ represents the
time of interest which in our case will be the ``observation
time''.

Inserting the equation (\ref{eq:d_k}) in the last
expression, we can rewrite
$\langle\delta\hat\phi'_k\rangle_\Omega$ as
\begin{widetext}
  \begin{multline}
    \langle\delta\hat\phi'_k\rangle_\Omega =
    \frac{2}{a(\eta_c)}\cdot\frac{k\eta_c\sqrt{\hbar L^3
        k}}{\left(1+10k^2\eta_c^2+9k^4\eta_c^4\right)^{1/4}}\cdot
    \frac{ x^R_k +i x_k^I
    }{\sqrt{1+5k^2\eta_c^2-\sqrt{1+10k^2\eta_c^2+9k^4\eta_c^4}}} 츾충 \\
    \Bigg\{\cos\Delta_k
    \sqrt{\sqrt{1+10k^2\eta_c^2+9k^4\eta_c^4} - 1
      +3k^2\eta_c^2 } \quad +    \\
    \sin\Delta_k
    \Bigg[\sqrt{\sqrt{1+10k^2\eta_c^2+9k^4\eta_c^4} + 1-
      3k^2\eta_c^2} - \\ \frac{1}{k\eta_c}
    \sqrt{\sqrt{1+10k^2\eta_c^2+9k^4\eta_c^4} - 1
      +3k^2\eta_c^2}\quad \Bigg] \Bigg\}. \\
  \end{multline}
\end{widetext}
Now we take the ensemble mean value of the square of
$\langle\delta\hat\phi'_k\rangle_\Omega$, taking out a
factor of $\kappa$ (remember that $\kappa = \hbar L^3 k/4
a^2$, see last section) and call it $C_{wigner}(k)$
\begin{multline}\label{C_wigner_2}
  C_{wigner}(k) = \frac{32 z_k^2}{\sqrt{1+10z_k^2+9z_k^4}}
  \times \\
  \frac{1}{1+5z_k^2-\sqrt{1+10z_k^2+9z_k^4}} \\
  \Bigg\{ \left[\sqrt{1+10z_k^2+9z_k^4} - 1
    +3z_k^2\right]\left(\cos\Delta_k
    -\frac{\sin\Delta_k}{z_k}\right)^2 +    \\
  \sin^2\Delta_k \left[\sqrt{1+10z_k^2+9z_k^4} - 3z_k^2 - 7
  \right] + \\ 8z_k\cos\Delta_k \sin\Delta_k \Bigg\},
\end{multline}
where we replaced $k\eta_k^c(k)$ by $z_k$. Henceforth
\eqref{eq:contacto_obs} is
\begin{equation}
  |\alpha_{lm}|^2_{ML}  = \frac{s^2\hbar}{2 \pi a^2}
  \int   \frac{C_{wigner}(x/R_D)}{x}\mathcal{T}(x/R_D)^2
  j_l^2(x)  dx.
\end{equation}

Now we are prepared to compare the predictions of the
various schemes of collapse with observations.

Before doing so it is worth recalling that the standard
results are obtained if the function $C$ is a constant, and
to mention that it turns out that in order to obtain a
constant $C$ (in this and any collapse scheme) there seems
to be a single simple option: That the $z_k$ be essentially
independent of $k$ indicating that the time of collapse for
the mode $k$, $\eta_k^c$ should depends of the mode
frequency according to $\eta_k^c = z/k$. For a more detailed
treatment we refer to the article \cite{Perez2006}.

\section{\label{sec:comp_observations}Comparing with
  Observations}

This is going to be a rather preliminary analysis
concentrating on the main features of the resulting spectrum
and ignoring the late time physics corresponding to the
effects of reheating and acoustic oscillations (represented
by $\mathcal{T}(k)$). Actual comparison with empirical data
requires a more involved analysis which is well outside the
scope of the present paper.

We remind the reader that $C(k)$ encapsulates all the
imprint of the details of the collapse scheme on the
observational power spectrum.

\begin{figure}
  \centering \subfigure[$C_1$, the two field variables
  $\langle \hat y_k\rangle$ and $\langle\hat\pi_k\rangle$,
  collapses to a random value of the dispersion of the
  vaccum state independently] {
    \label{fig:c_1}
    \resizebox{70mm}{!}{\includegraphics{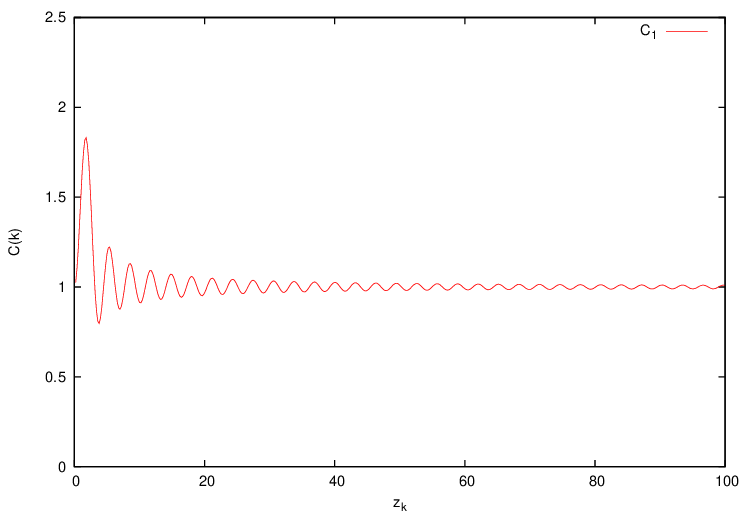}} }
  \hspace{1cm} \subfigure[$C_2$, this scheme is proposed
  taking in account the fact that only
  $\langle\hat\pi_k\rangle$ appears in the EFE at first
  order.]{
    \label{fig:c_2}
    \resizebox{70mm}{!}{\includegraphics{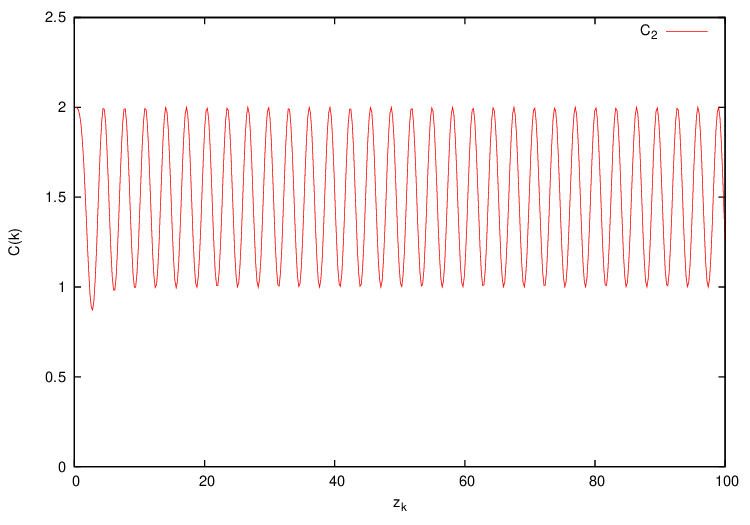}} }
  \hspace{1cm} \subfigure[$C_{wigner}$, this scheme proposes
  a kind of correlation between the post-collapse values
  taking the Wigner functional of the vaccum state as an
  indicator of this correlation.]{
    \label{fig:c_wigner}
    \resizebox{70mm}{!}{\includegraphics{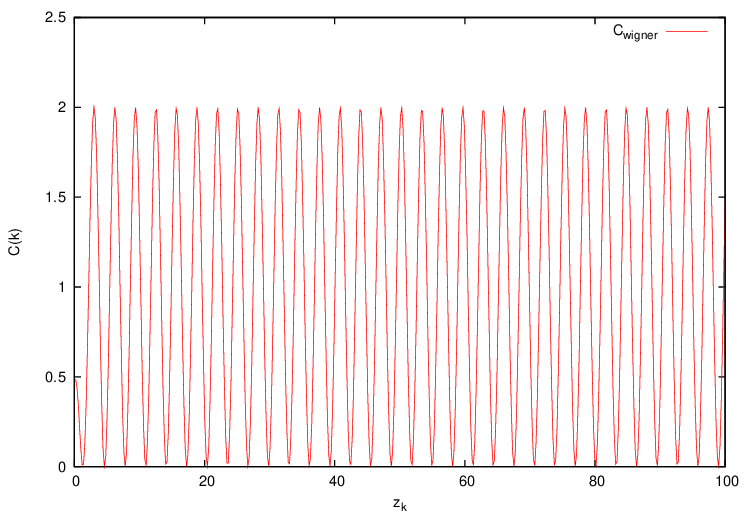}} }
  \caption{Plots of the three collapse schemes, we could
    apreciate that $C_2$ (middle) and $C_{wigner}$ have a
    similar behavior despite their dissimilar functional
    form.}
  \label{fig:c_k}
\end{figure}

The functional form of this quantity for the scheme
considered in this article, $C_{wigner}$ \eqref{C_wigner_2},
has a more complicated form than the corresponding
quantities that resulted from the schemes of collapse
considered in \cite{Perez2006}. Here we reproduce those
expressions for comparison with the scheme considered here
and with observations.
In the first collapse scheme \eqref{C_sudarsky}, the
expectation values for the field $\hat y_k$ and its
canonical conjugate momentum $\hat\pi_k$ after the collapse
are randomly distributed within the respective ranges of
uncertainties in the pre-collapsed state, and are
uncorrelated. The resulting power spectrum has
\begin{equation}\label{C_sudarsky}
  C_1(k) = 1 + \frac{2}{z_k^2}\sin^2\Delta_k + \frac{1}{z_k}\sin(2\Delta_k).
\end{equation}
The second scheme considered in \cite{Perez2006} only the
conjugate momentum changes its expectation value from zero
to a value in such range, this second scheme is proposed
since in the first-order equation
(\ref{eq:semiclassical-fundamental}) only this variable
appears as a source. This leads to a spectrum with
\begin{equation}\label{C_2}
  C_2(k) = 1 + \sin^2\Delta_k \left(1 - \frac{1}{z_k^2}\right)
  - \frac{1}{z_k}\sin (2\Delta_k).
\end{equation}
Despite the fact that the expression for $C_{wigner}$ looks
by far more complicated that $C_2$, their dependence in
$z_k$ is very similar, except for the amplitude of the
oscillations (see figures \ref{fig:c_2} and
\ref{fig:c_wigner}).  Another interesting fact that can be
easily detected in the behaviour of the different schemes of
collapse is that if we consider the limit $z_k \to \pm
\infty$, then $C_1(k) \to 1$ and we recover the standard
scale invariant spectrum. This does not happen with $C_2(k)$
or $C_{wigner}(k)$ (see figure \ref{fig:c_k}).

We recall that the standard form of the predicted spectrum
is recovered by taking $C(k) = 1$. Therefore, we can
consider the issue of how the various collapse schemes
approach the standard answer (given the fact that the
standard answer seems to fit the observations rather
well). In particular we want to investigate how sensitive
are the predictions for the various schemes, to small
departures from the case where $z_k$ is independent of $k$,
which as we argued above would lead to a precise agreement
with the standard spectral form. In order to carry out this
analysis, we must obtain the integrals
\eqref{eq:contacto_obs} for the various collapse schemes
characterized by the various functions $C_1(k), C_2(k)$ and
$C_{wigner}(k)$. It is convenient to define the adimensional
quantity $\tilde{z}_x \equiv x N(x)$, where $x= k R_D$ and
$N(x) \equiv \eta_{k(x)}^c/R_D$. We will be working under
the following assumptions: (1) The changes in scale during
the time elapsed from the collapse to the end of inflation
are much more significant than those associated the time
elapsed from the end of inflation to our days, thus we will
use the approximation $\Delta_k = -\tilde{z_x}$; (2) We will
explore the sensitivity for small deviations of the ``$z_k$
independent of $k$ recipe'' by considering a linear
departure from the $k$ independent $z_k$ characterized by
$\tilde{z}_x$ as $\tilde{z}_x = A + Bx$ in order to examine
the robustness of the collapse scheme in predicting the
standard spectrum.  We note that $A$ and $B$ are
adimensional.

\begin{figure*}
  \includegraphics[width=180mm, angle=90]{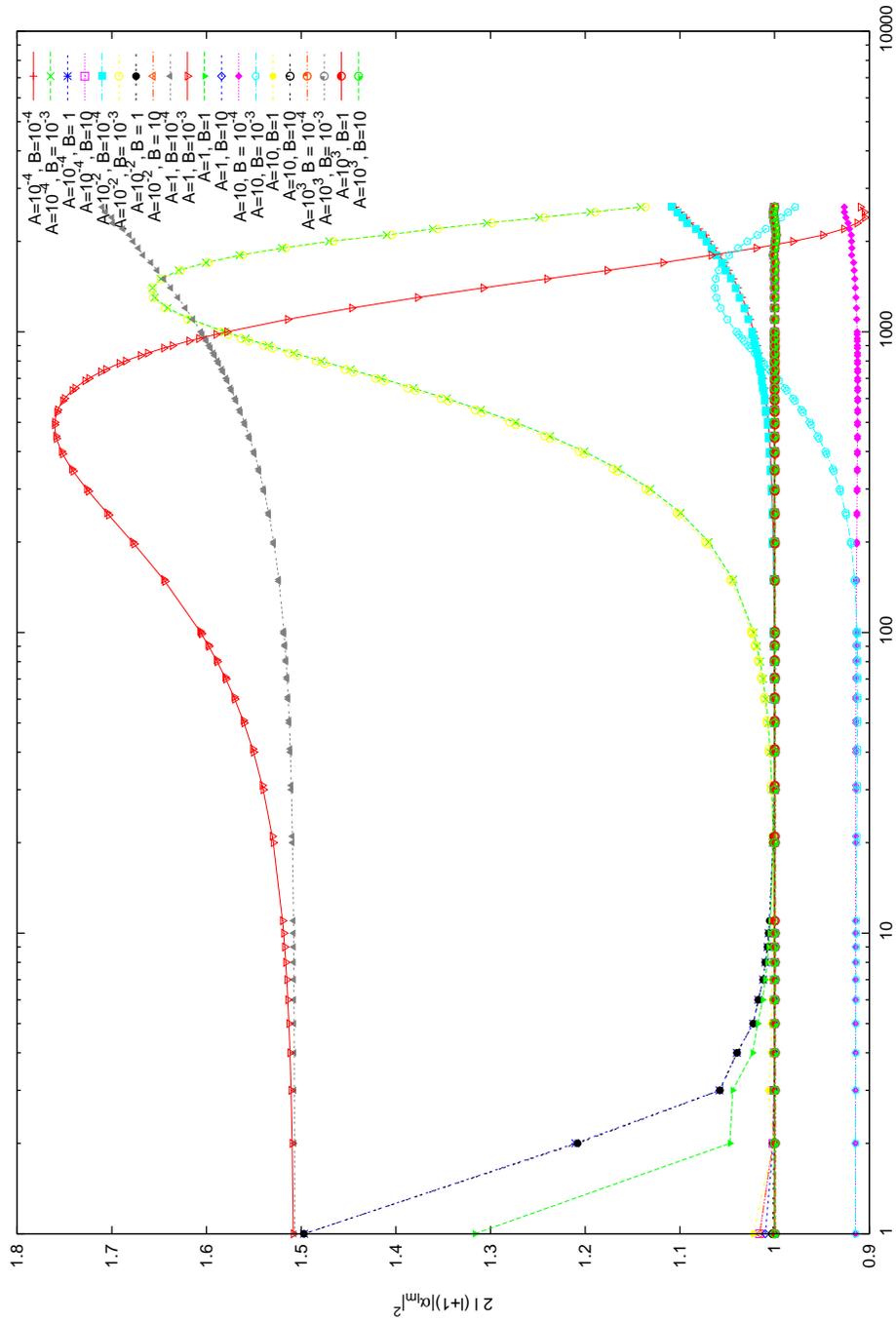}
  \caption{Semilog plot of $|\alpha_{lm}|^2(C_1(k))$ for
    different values of $(A,B)$, representing how robust is
    the scheme of collapse when it departs from $z_k$
    constant. The abscissa is $l$ until $l=2600$}
  \label{fig:c1_log}
\end{figure*}

\begin{figure*}
  \includegraphics[width=180mm, angle=90]{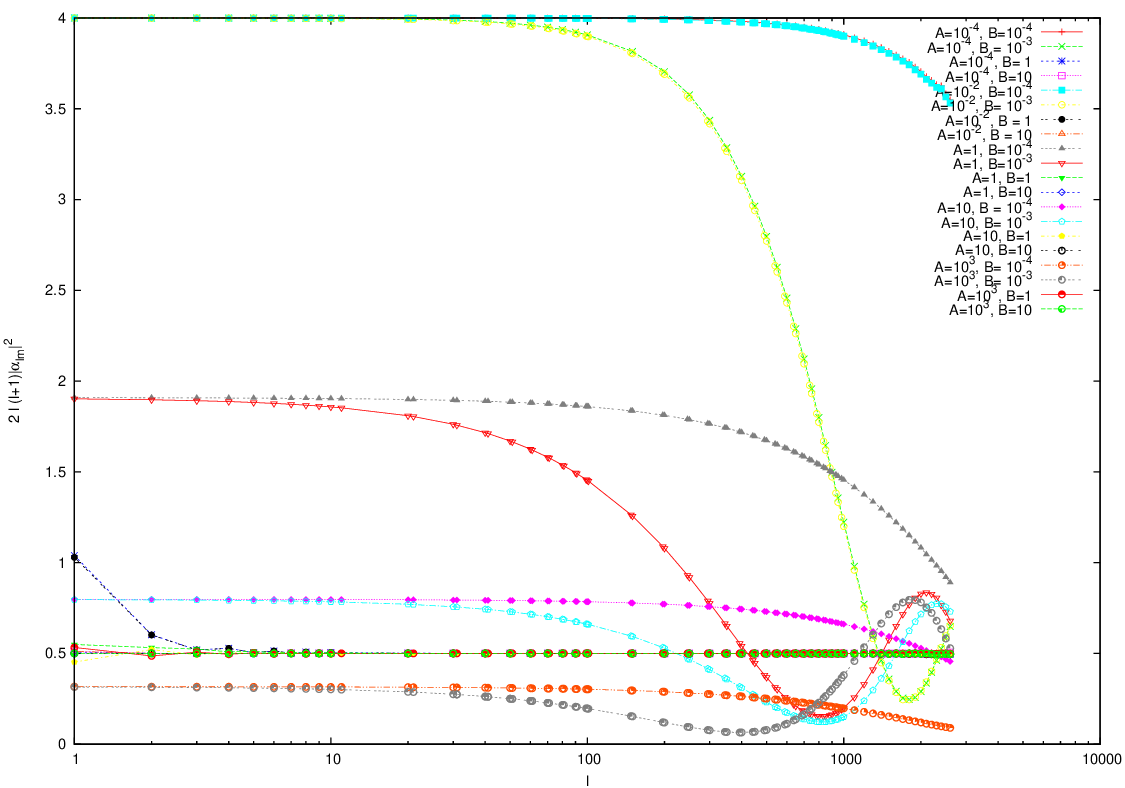}
  \caption{Semilog plot of $|\alpha_{lm}|^2(C_2(k))$ for
    different values of $(A,B)$, representing how robust is
    the scheme of collapse when it departs from $z_k$
    constant.The abscissa is $l$ until $l=2600$}
  \label{fig:c2_log}
\end{figure*}

\begin{figure*}
  \includegraphics[width=180mm, angle=90]{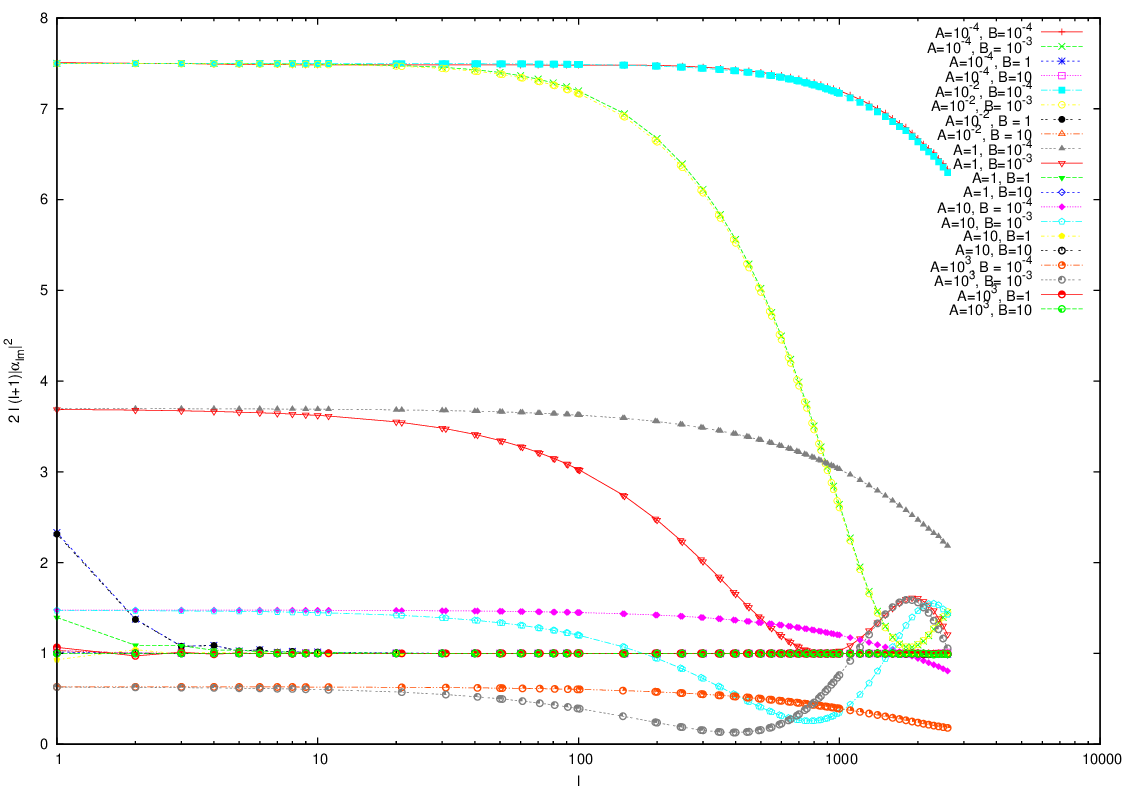}
  \caption{Semilog plot of $|\alpha_{lm}|^2(C_{wigner}(k))$
    for different values of $(A,B)$, representing how robust
    is the scheme of collapse when it departs from $z_k$
    constant. The abscissa is $l$ until $l=2600$.}
  \label{fig:c_wigner_log}
\end{figure*}

\begin{figure*}
  \centering \mbox { \subfigure[$C_{1}(k),\, A = 10^{-4}$]{
      \includegraphics[width=65mm,
      height=50mm]{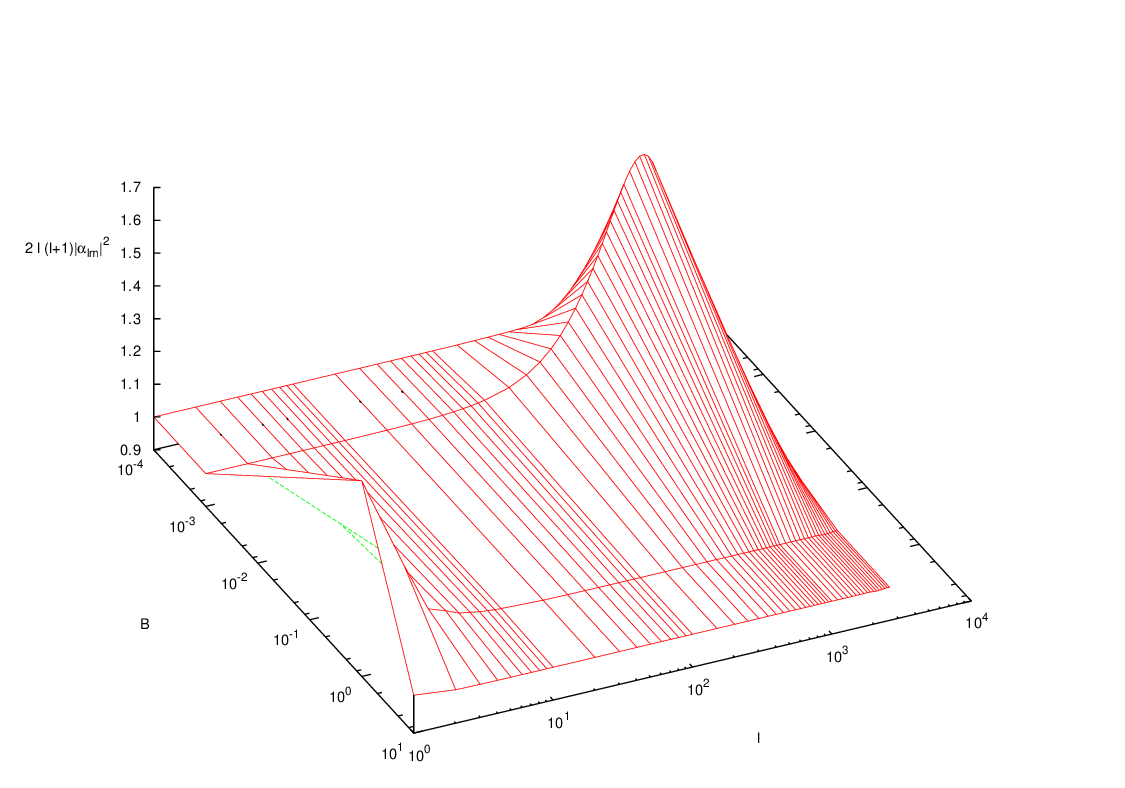}
      \label{fig:c1_a_10-4}} \subfigure[$C_{1}(k),\, A =
    10^{-3}$]{
      \includegraphics[width=65mm,
      height=50mm]{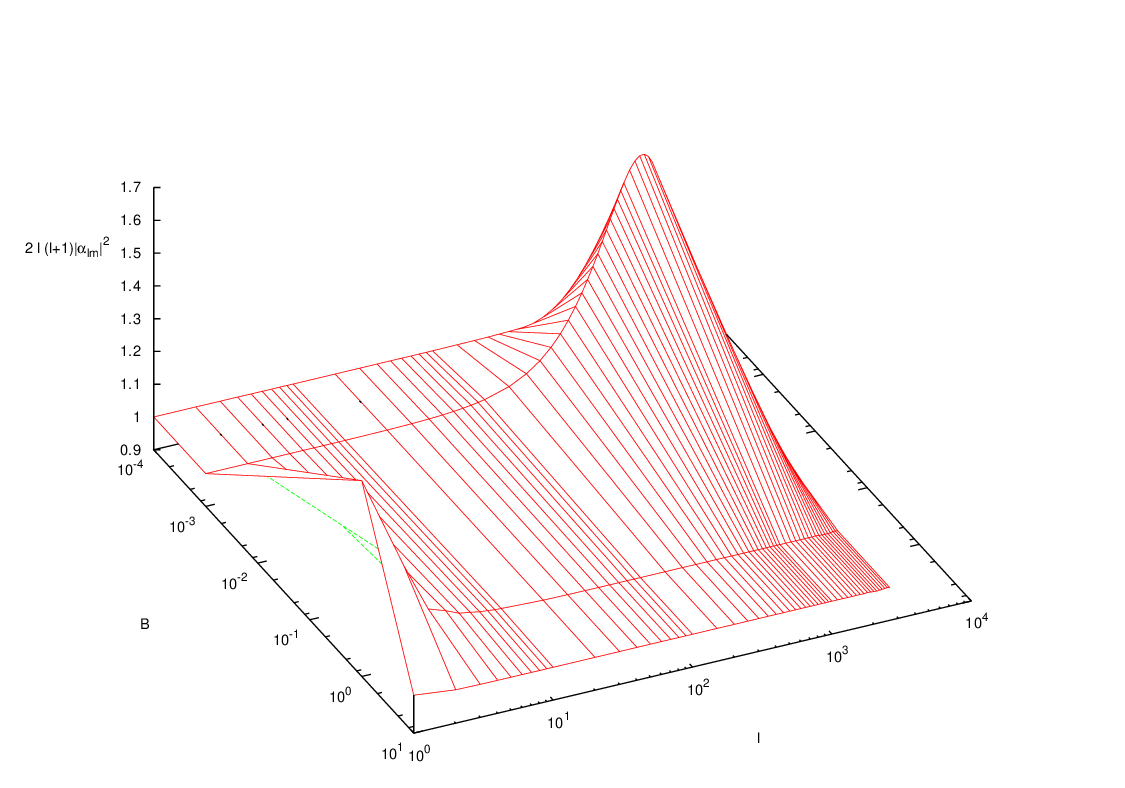}
      \label{fig:c1_a_10-2}} } \mbox {
    \subfigure[$C_{1}(k),\, A = 1$]{
      \includegraphics[width=65mm,
      height=50mm]{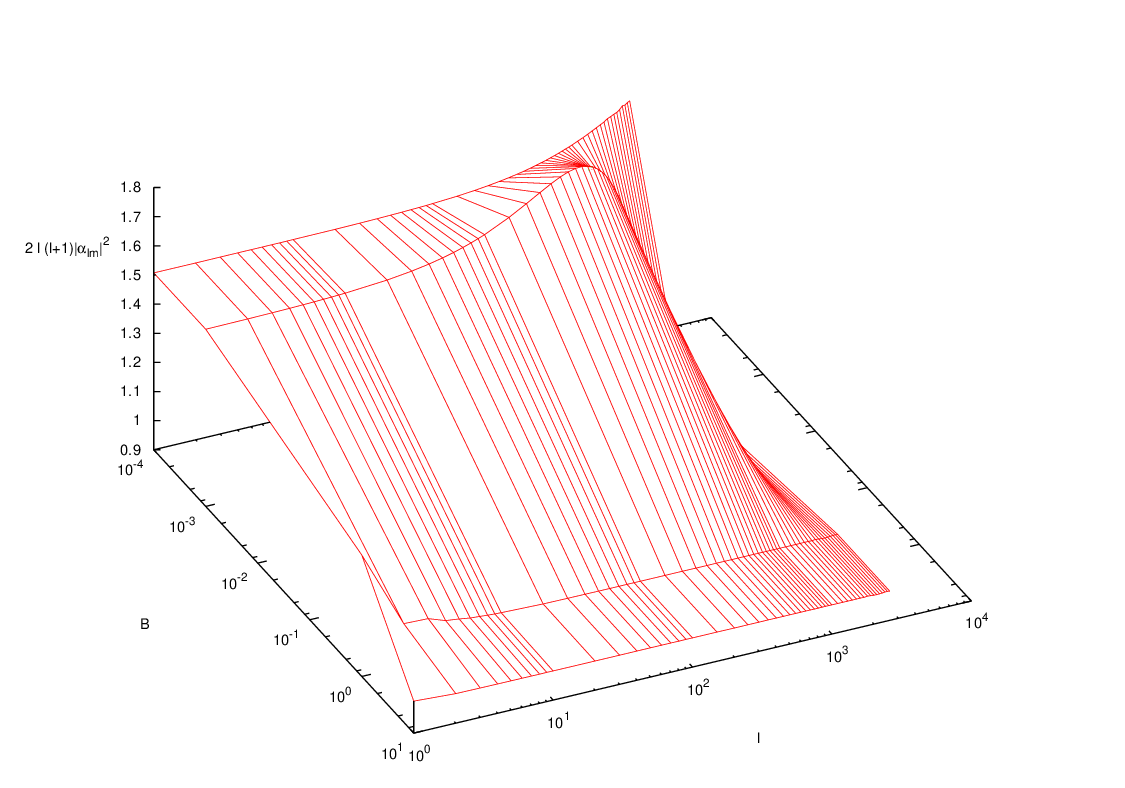}
      \label{fig:c1_a_1}}

    \subfigure[$C_{1}(k),\, A = 10$]{
      \includegraphics[width=65mm,
      height=50mm]{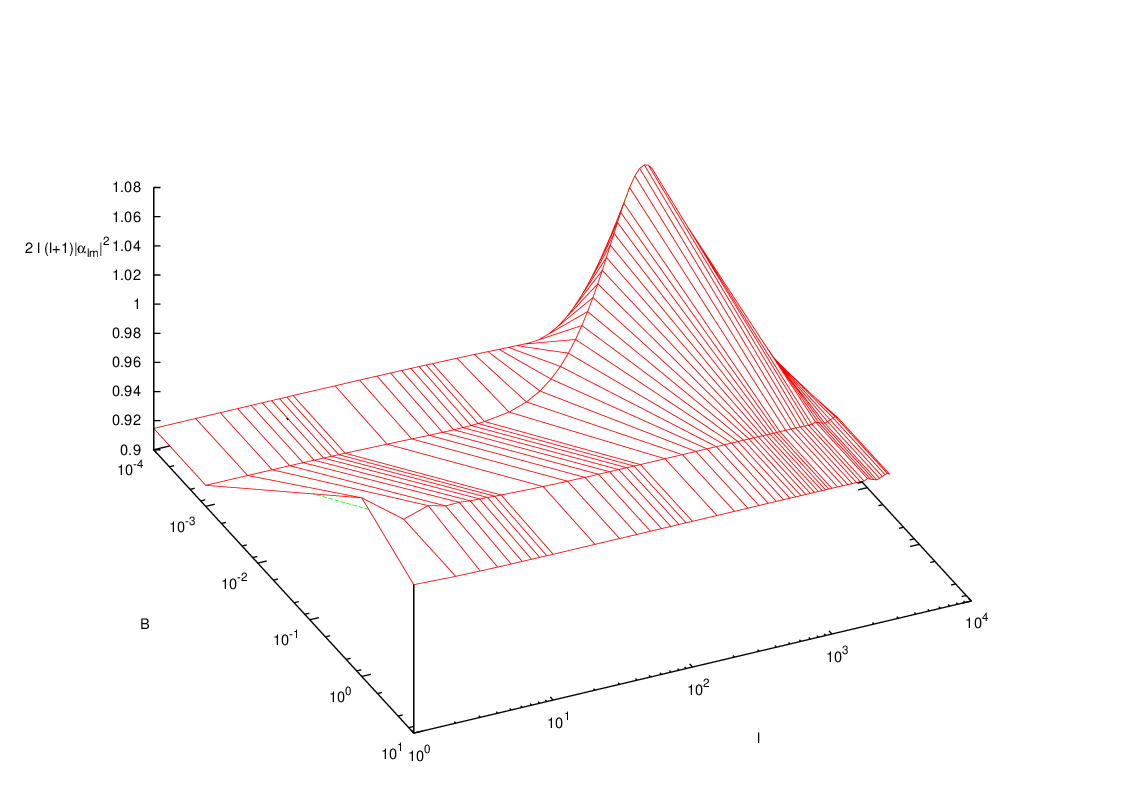}
      \label{fig:c1_a_10}
    } } \subfigure[$C_{1}(k),\, A = 1000$]{
    \includegraphics[width=65mm,
    height=50mm]{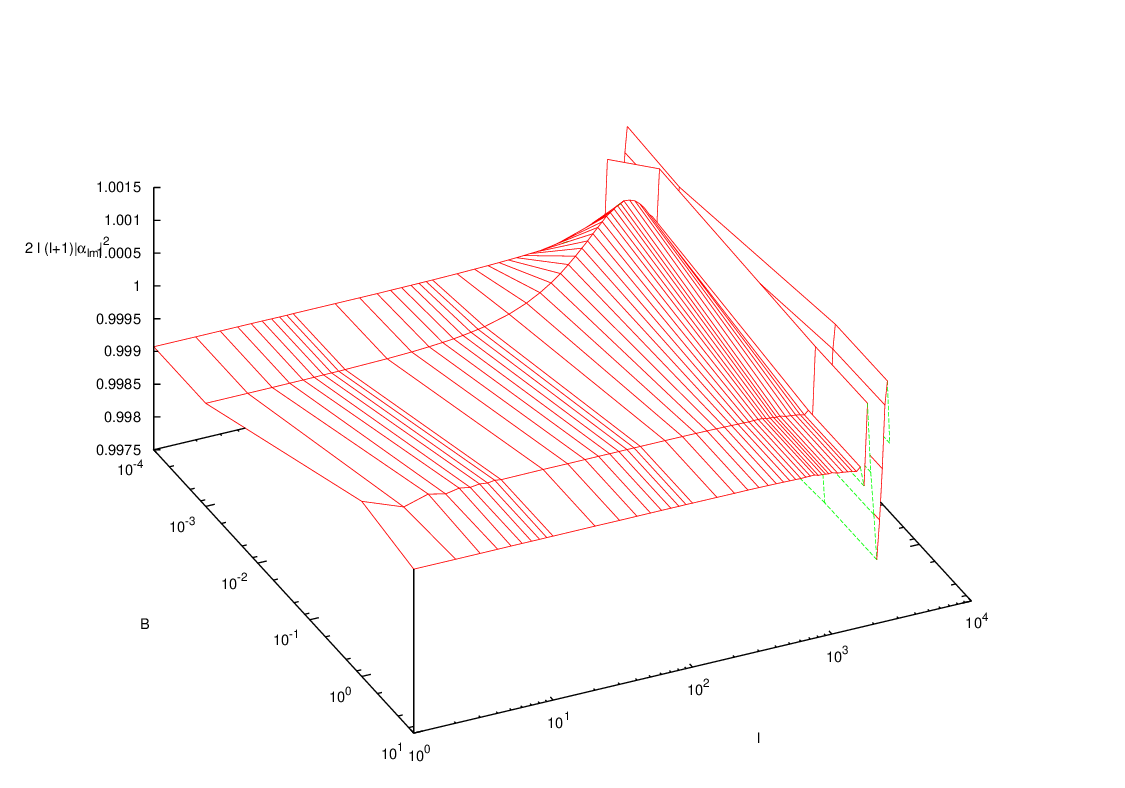}
    \label{fig:c1_a_1000}
  }

  \caption{Plot showing how the integral of
    $|\alpha_{lm}|^2(C_{1})$ varies respect changes in $B$
    ($10^{-4} - 10$), keeping $A$ fixed. Both axes $B$ and
    $l$ are in logscale. See the main text for a more
    extensive explanation.}  \label{fig:c1_3d}
\end{figure*}

\begin{figure*}

  \centering \mbox { \subfigure[$C_{2}(k),\, A = 10^{-4}$]{
      \includegraphics[width=65mm,
      height=50mm]{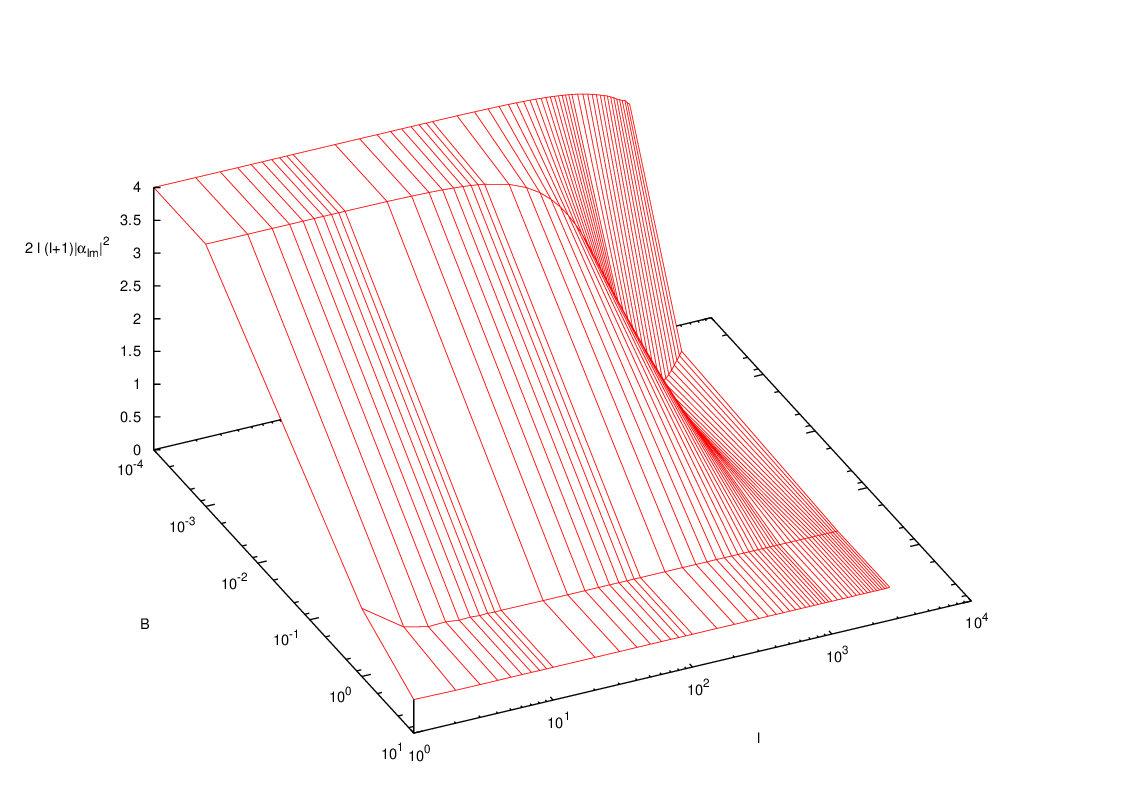}
      \label{fig:c2_a_10-4}} \subfigure[$C_{2}(k),\, A =
    10^{-3}$]{
      \includegraphics[width=65mm,
      height=50mm]{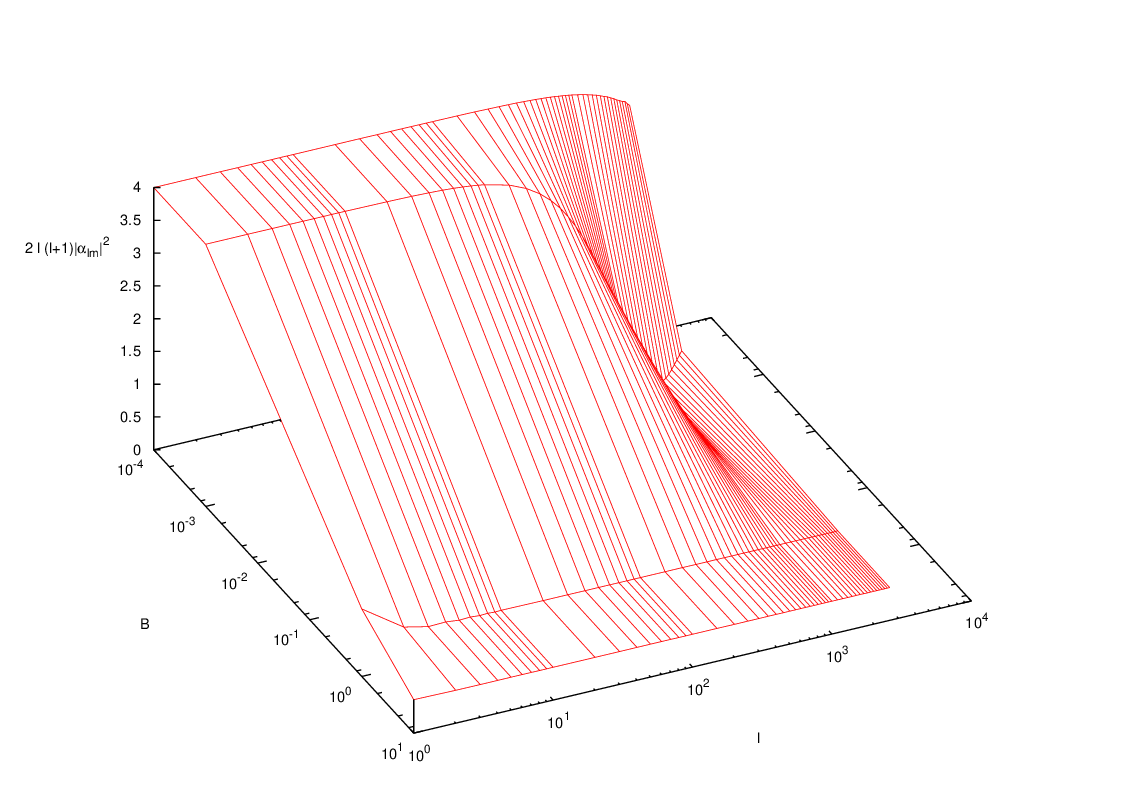}
      \label{fig:c2_a_10-2}} } \mbox {
    \subfigure[$C_{2}(k),\, A = 1$]{
      \includegraphics[width=65mm,
      height=50mm]{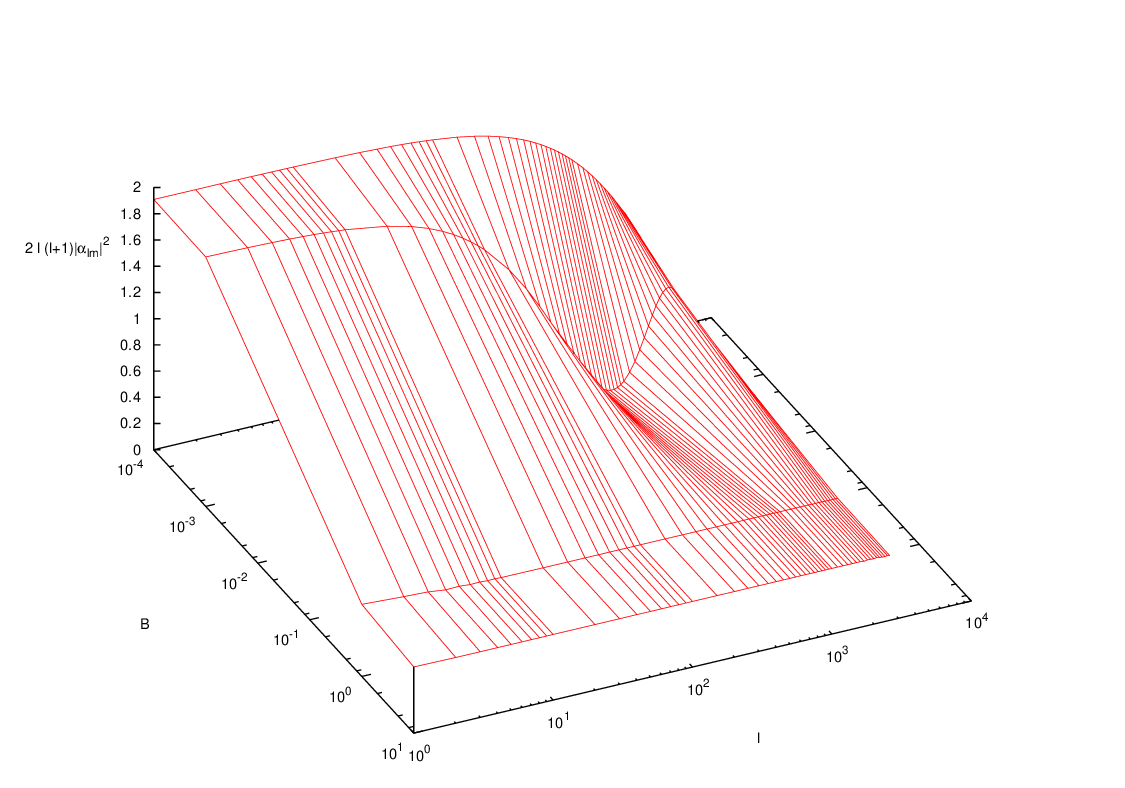}
      \label{fig:c2_a_1}}

    \subfigure[$C_{2}(k),\, A = 10$]{
      \includegraphics[width=65mm,
      height=50mm]{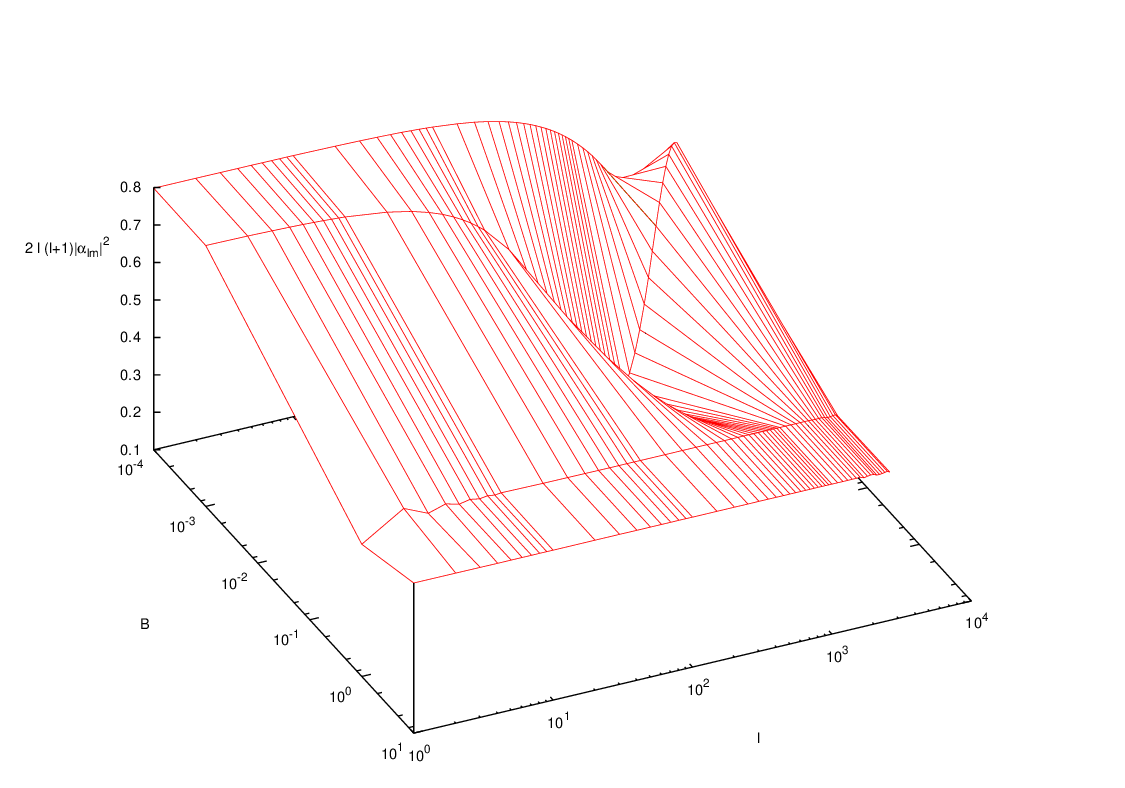}
      \label{fig:c2_a_10}
    } } \subfigure[$C_{2}(k),\, A = 1000$]{
    \includegraphics[width=65mm,
    height=50mm]{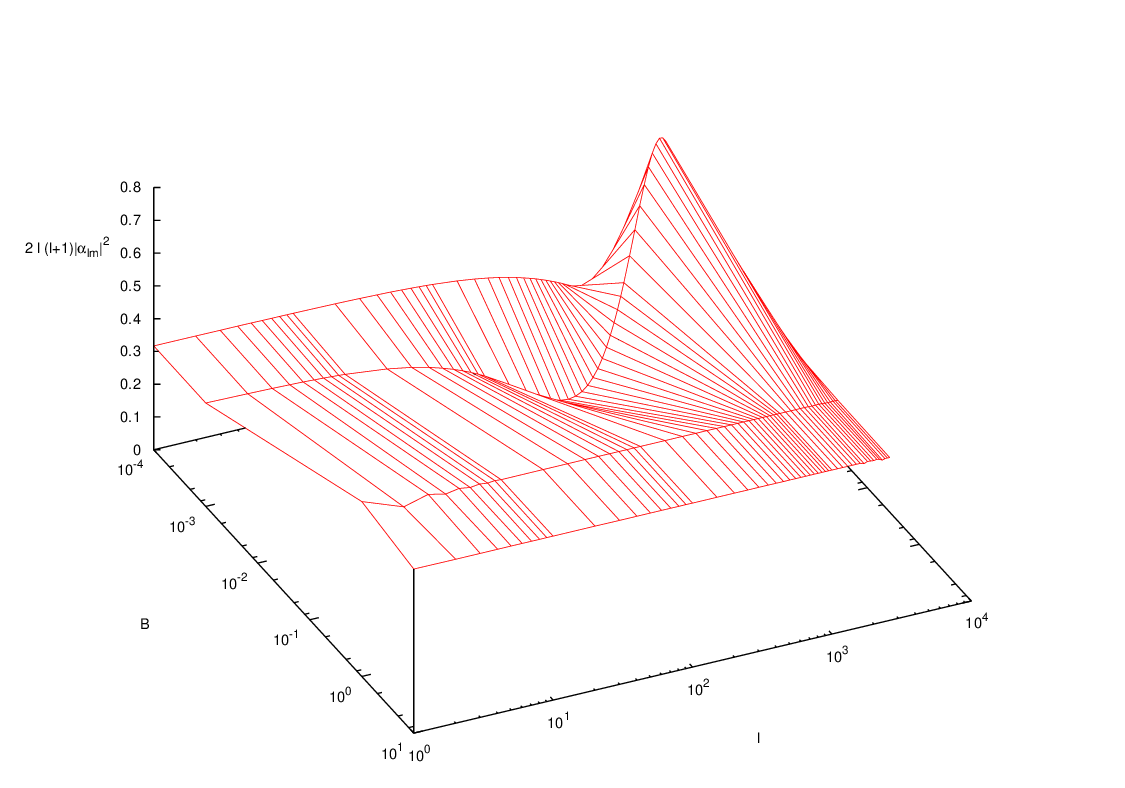}
    \label{fig:c2_a_1000}
  }

  \caption{Plot showing how the integral of
    $|\alpha_{lm}|^2(C_{2})$ varies respect changes in $B$
    ($10^{-4} - 10$), keeping $A$ fixed. Both axes $B$ and
    $l$ are in logscale. See the main text for a more
    extensive explanation.}  \label{fig:c2_3d}
\end{figure*}

\begin{figure*}

  \centering \mbox { \subfigure[$C_{wigner}(k),\, A =
    10^{-4}$]{
      \includegraphics[width=65mm,
      height=50mm]{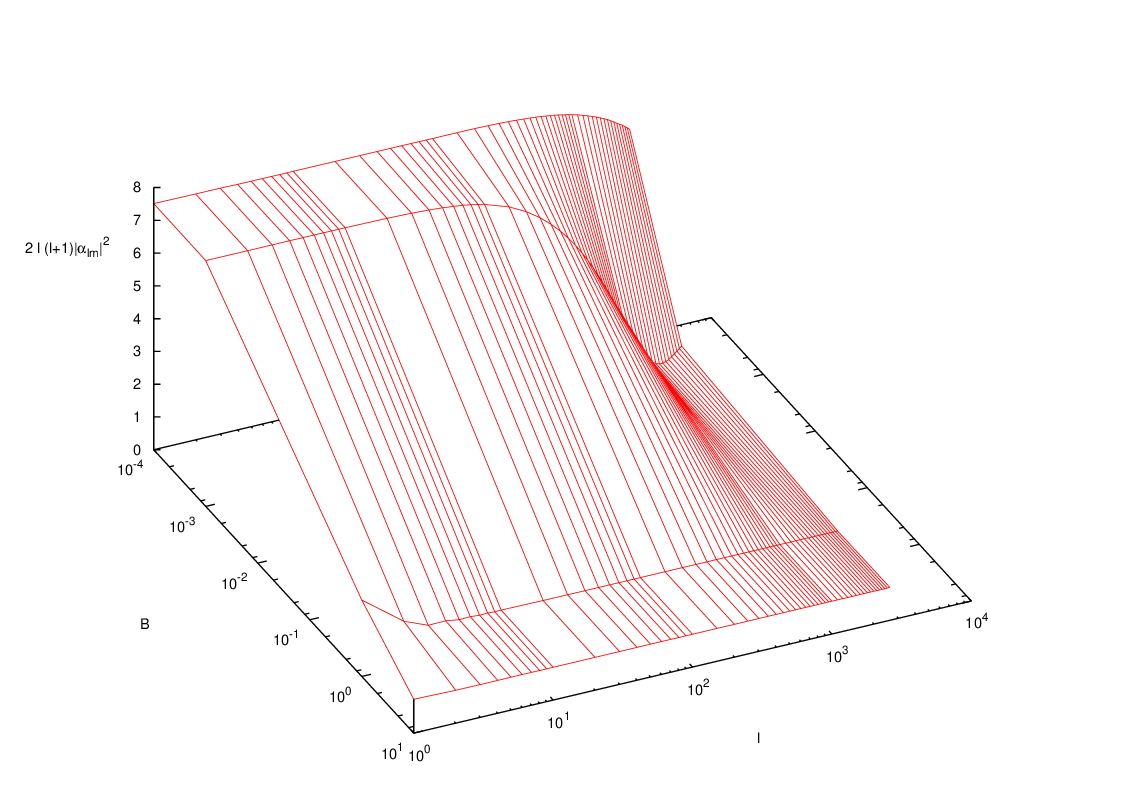}
      \label{fig:wigner_a_10-4}}
    \subfigure[$C_{wigner}(k),\, A = 10^{-3}$]{
      \includegraphics[width=65mm,
      height=50mm]{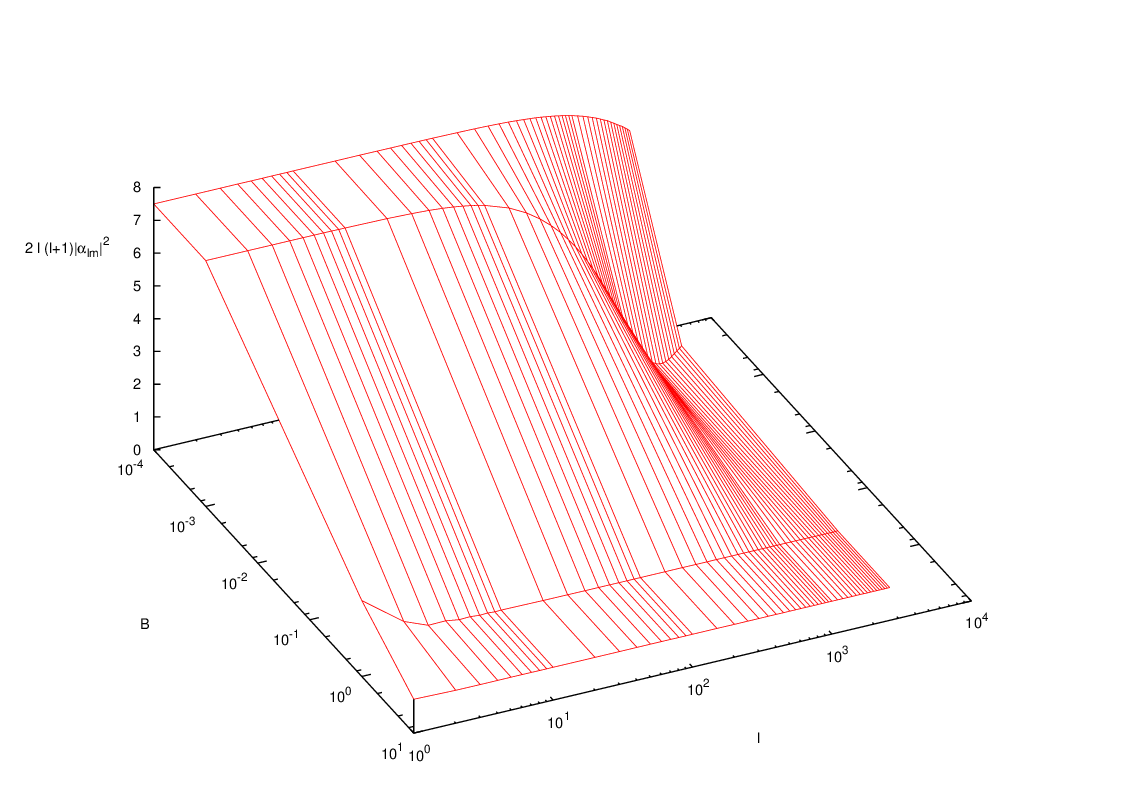}
      \label{fig:wigner_a_10-2}} } \mbox {
    \subfigure[$C_{wigner}(k),\, A = 1$]{
      \includegraphics[width=65mm,
      height=50mm]{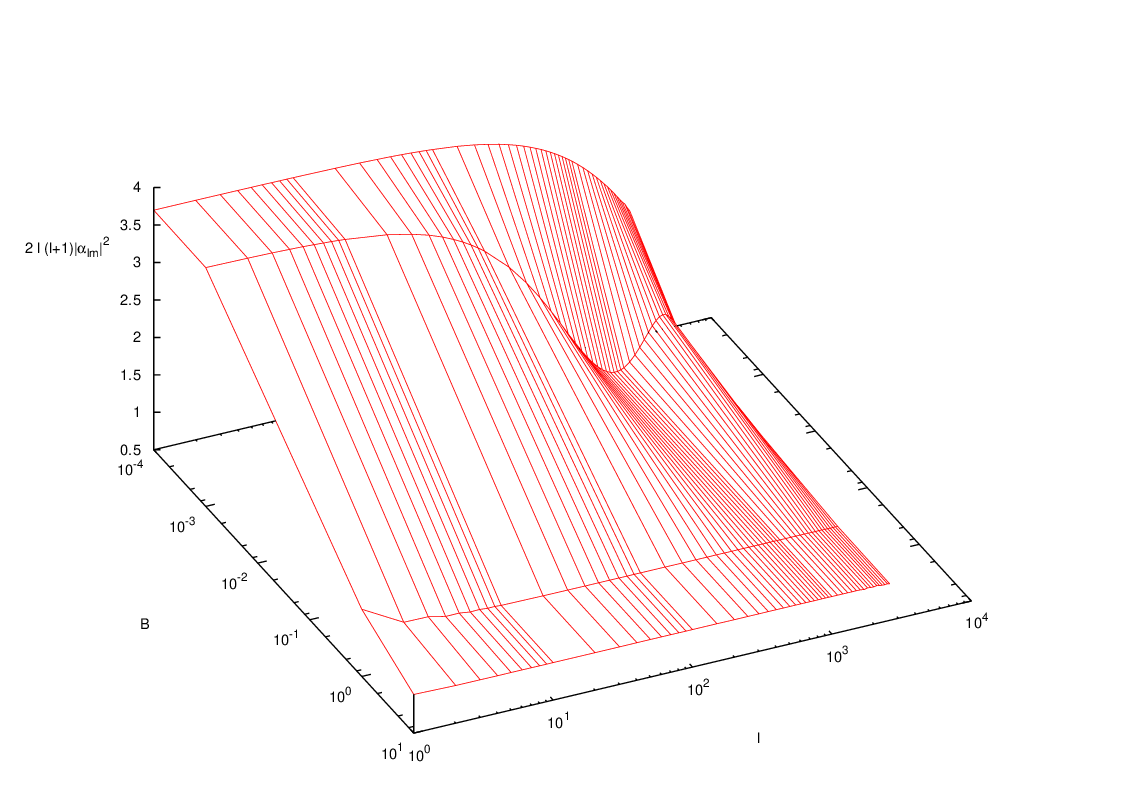}
      \label{fig:wigner_a_1}}

    \subfigure[$C_{wigner}(k),\, A = 10$]{
      \includegraphics[width=65mm,
      height=50mm]{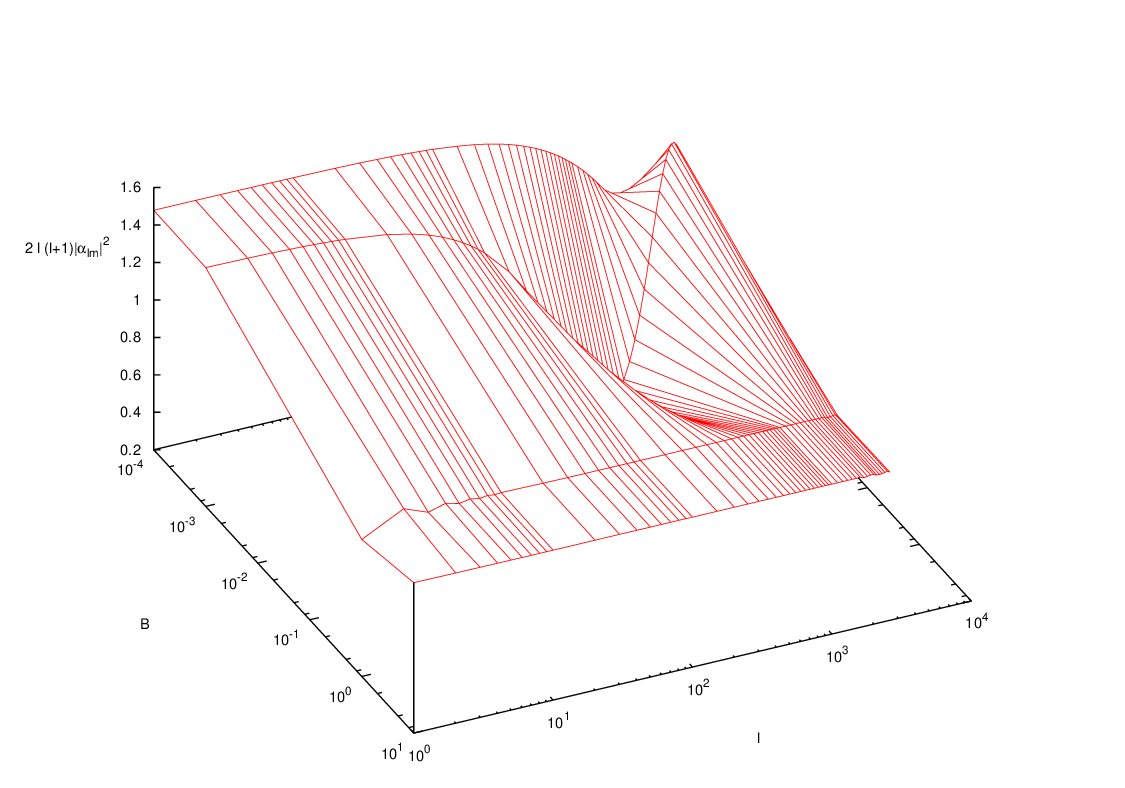}
      \label{fig:wigner_a_10}
    } } \subfigure[$C_{wigner}(k),\, a = 1000$]{
    \includegraphics[width=65mm,
    height=50mm]{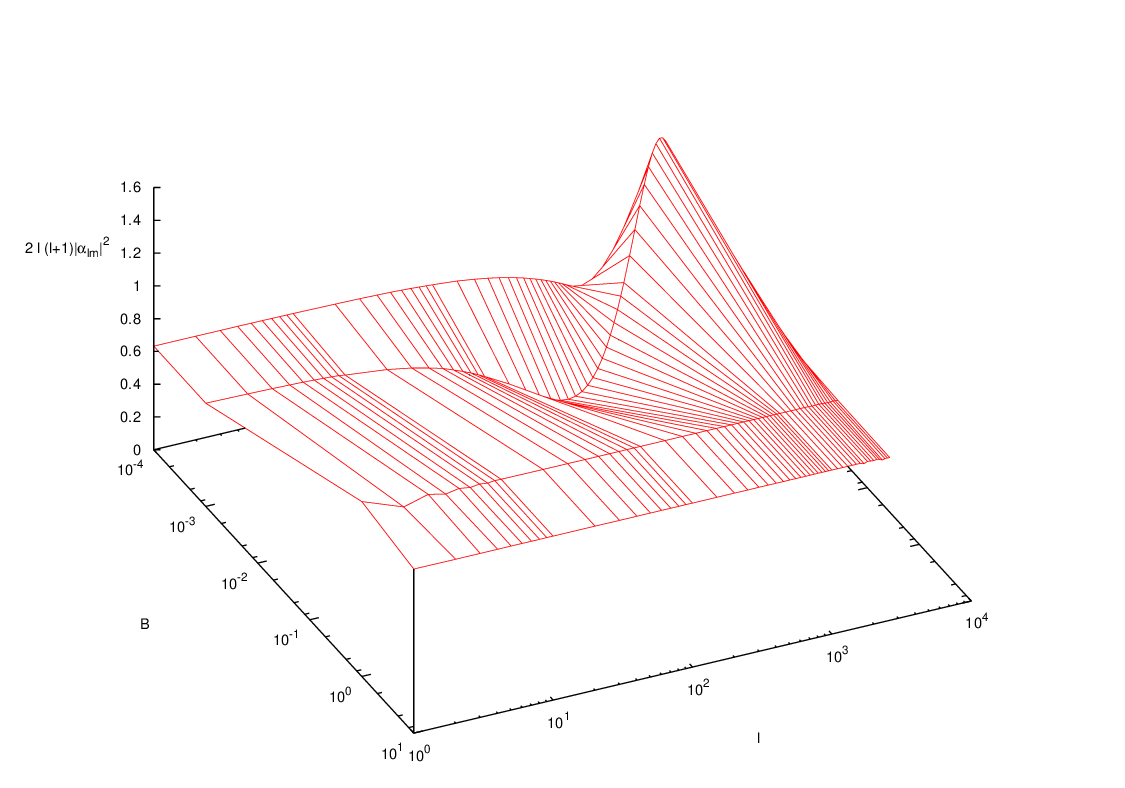}
    \label{fig:wigner_a_1000}
  }

  \caption{Plot showing how the integral of
    $|\alpha_{lm}|^2(C_{wigner})$ varies respect changes in
    $B$ ($10^{-4} - 10$), keeping $A$ fixed. Both axes $B$
    and $l$ are in logscale. See the main text for a more
    extensive explanation.}  \label{fig:c_wigner_3d}
\end{figure*}

In the figures \ref{fig:c1_log}, \ref{fig:c2_log} and
\ref{fig:c_wigner_log} reflect the way the spectrum behaves
as a function of $\, l\,$, were we must recall that standard
prediction (ignoring the late physics input of plasma
oscillations etc) is a horizontal line.  Those graphs
represent various values of $A$ and $B$ chosen to sample a
relatively ample domain.  The graphs (\ref{fig:c1_3d},
\ref{fig:c2_3d} and \ref{fig:c_wigner_3d}) show the form of
the spectrum for various choices for the value of $B$
keeping the value of $A$ fixed.

It is important at this point remind the reader -in the
order to avoid possible misinterpretations- that this graphs
are ignoring the effect of late physics phenomena (plasma
oscillations, etc.). Our aim, at this stage is to compare
this graphs with the scale-invariant spectrum predicted by
standard inflationary scenarios (i.e. a constant value for
$2l(l+1)|\alpha_{lm}|^2$) and not -directly- with the
observed spectrum.

As we observed before the behavior of $C_2$ and $C_{wigner}$
is qualitatively similar, the main difference comes from the
amplitude of the oscillations of the functional.


From these results we can obtain some reasonable constrains
on the values of the $A$ and $B$ for the different schemes
of collapse.  We start by defining for a given predicted
spectrum the degree of deviation from the flat spectrum to
be simply $\Delta_{lmax} \equiv (\frac{1}{lmax}
\Sigma_{l=1}^{l=lmax} [(l(l+1) \frac{1}{2l+1}$ $\Sigma_m
|\alpha_{l,m} |^2 -S]^2 )^{1/2} /S $ where $S$ represents
the flat spectrum that would best approximate the
corresponding imaginary data and is given by $S\equiv
\frac{1}{lmax} \Sigma_{l=1}^{l=lmax} (l(l+1) \frac{1}{2l+1}
\Sigma_m |\alpha_{l,m} |^2)$. If we set a bound on the
departure from scale invariance up to $l=1500$ of 10\%
measured by $\Delta_{lmax}$ (i.e. requiring $\Delta_{lmax}
<0.1$) we obtain for the various collapse schemes the
corresponding allowed range of values for the parameters $A$
and $B$. The results from these analysis are presented in
the tables \ref{tab:c_1}, \ref{tab:c_2} and
\ref{tab:c_wigner}.  We see that the restriction of range in
$B$ becomes weaker for larger values of $A$, something that
can be described by stating that the earlier the collapse
occurs the larger the possible departures from the behavior
$\eta^c_k k = constant$.

\begin{table}
  \caption{\label{tab:c_1}Robustness of $C_{1}$ when the parameters
    $(A,B)$ were varied from $10^{-4} \leq A \leq 10^3$ and
    $10^{-4} \leq B \leq 10$}
  \begin{ruledtabular}
    \begin{tabular}{ccc}
      \multicolumn{3}{|c|}{$C_1(k)$}\\
      \hline
      $A$&$B$&$\Delta_{lmax} \times 100$ \\
      \hline
      0.0001    &   0.0001      &       6.63019  \\
      0.0001    &    0.001      &       28.3844  \\
      0.0001    &        1      &      0.288273  \\
      0.0001    &       10      &      0.301883  \\
      0.01    &   0.0001      &       6.84475  \\
      0.01    &    0.001      &       28.3706  \\
      0.01    &        1      &      0.282546  \\
      0.01    &       10      &      0.301614  \\
      1    &   0.0001      &       10.1258  \\
      1    &    0.001      &       21.3117  \\
      1    &        1      &      0.247444  \\
      1    &       10      &      0.341509  \\
      10    &   0.0001      &       1.67782  \\
      10    &    0.001      &       15.8869  \\
      10    &        1      &      0.195523  \\
      10    &       10      &      0.384265  \\
      1000    &   0.0001      &       0.44236  \\
      1000    &    0.001      &       1.58567  \\
      1000    &        1      &      0.394892  \\
      1000    &       10      &      0.402706  \\
    \end{tabular}
  \end{ruledtabular}
\end{table}

\begin{table}
  \caption{  \label{tab:c_2}
    Robustness of $C_{2}$ when the parameters
    $(A,B)$ were varied from $10^{-4} \leq A \leq 10^3$ and
    $10^{-4} \leq B \leq 10$}
  \begin{ruledtabular}
    \begin{tabular}{ccc}
      \multicolumn{3}{|c|}{$C_2(k)$}\\
      \hline
      $A$&$B$&$\Delta_{lmax} \times 100$ \\
      \hline
      0.0001    &   0.0001      &       7.92849  \\
      0.0001    &    0.001      &       53.9872  \\
      0.0001    &        1      &      0.423473  \\
      0.0001    &       10      &      0.249129  \\
      0.01    &   0.0001      &       8.12093  \\
      0.01    &    0.001      &       54.2265  \\
      0.01    &        1      &      0.277929  \\
      0.01    &       10      &      0.251313  \\
      1    &   0.0001      &       21.8266  \\
      1    &    0.001      &       50.6328  \\
      1    &        1      &      0.312876  \\
      1    &       10      &      0.443572  \\
      10    &   0.0001      &       18.4953  \\
      10    &    0.001      &       46.1397  \\
      10    &        1      &      0.917963  \\
      10    &       10      &      0.445398  \\
      1000    &   0.0001      &       28.9085  \\
      1000    &    0.001      &       56.2369  \\
      1000    &        1      &      0.208227  \\
      1000    &       10      &      0.434914  \\
    \end{tabular}
  \end{ruledtabular}

\end{table}

\begin{table}
  \caption{\label{tab:c_wigner}Robustness of $C_{wigner}$ when the parameters
    $(A,B)$ were varied from $10^{-4} \leq A \leq 10^3$ and
    $10^{-4} \leq B \leq 10$}

  \begin{ruledtabular}
    \begin{tabular}{ccc}
      \multicolumn{3}{|c|}{$C_{wigner}(k)$}\\
      \hline
      $A$&$B$&$\Delta_{lmax} \times 100$ \\
      \hline
      0.0001    &   0.0001      &       10.0763  \\
      0.0001    &    0.001      &       47.3616  \\
      0.0001    &        1      &      0.506768  \\
      0.0001    &       10      &      0.162458  \\
      0.01    &   0.0001      &       10.2874  \\
      0.01    &    0.001      &        47.494  \\
      0.01    &        1      &      0.359852  \\
      0.01    &       10      &      0.165756  \\
      1    &   0.0001      &        18.445  \\
      1    &    0.001      &       34.1731  \\
      1    &        1      &      0.358535  \\
      1    &       10      &      0.394309  \\
      10    &   0.0001      &       19.3128  \\
      10    &    0.001      &       45.1946  \\
      10    &        1      &       0.51842  \\
      10    &       10      &      0.430548  \\
      1000    &   0.0001      &       28.9273  \\
      1000    &    0.001      &       56.2646  \\
      1000    &        1      &      0.197662  \\
      1000    &       10      &      0.445794  \\
    \end{tabular}
  \end{ruledtabular}
\end{table}

\begin{figure*}
  \centering \subfigure[$C_1$, with $A= 10, B = 1$.]{
    \label{fig:c1_tiempos_colapso}
    \includegraphics[width=90mm,
    height=53mm]{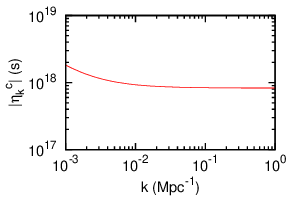}
  } \hspace{1cm} \subfigure[$C_2$, with $A=1000, B=1$.]{
    \label{fig:c2_tiempos_colapso}
    \includegraphics[width=90mm,
    height=53mm]{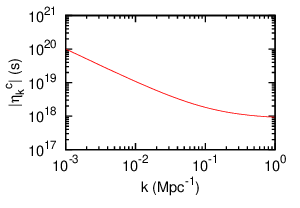}
  } \hspace{1cm} \subfigure[ $C_{wigner}$, with $A=0.01,
  B=10$. Note how in this scheme almost all the modes must
  collapse at the same time.]{
    \label{fig:cwigner_tiempos_colapso}
    \includegraphics[width=90mm,
    height=53mm]{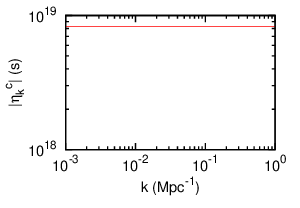}
  } \caption{Logarithmic plot in both axes of the collapse
    times $|\eta_k^c|$ (in seconds), for the three schemes,
    taking in account only the best values of $(A,B)$ in the
    range of $10^{-3}$ Mpc$^{-1}$ $< k < 1$ Mpc$^{-1}$. For
    these plots: $h = 0.7$.}
  \label{fig:tiempos_colapso}
\end{figure*}

We note that we can recover the range of times of collapse
for the different values of $A$ and $B$. We can solve $N(x)
= A/x + B$, therefore $|\eta_k^c(k)| = A/k + R_D B$. Note
that $R_D$ is the \emph{comoving} radii of the last
scattering surface. Considering the radial null geodesics we
find $R_{D} = \eta_0 - \eta_{d} $, where $\eta_d$ is the
time of the decoupling. The decoupling of photons occurs in
the matter domination epoch, so we can use the expression
for $R_D$ in terms of the scale factor, using the
corresponding solution to the Friedman equation
\begin{equation}
  R_D = \frac{2}{H_0}\left( 1 - \sqrt{a_d}\right),
\end{equation}
where we have normalized the scale factor so today is $a_0 =
1$, so, $a_d \equiv a(\eta_d) \simeq 10^{-3}$ and $H_0$ is
the Hubble variable today. The numerical value is $R_D =
5807.31 h^{-1}$ Mpc. Henceforth
\begin{equation}
  |\eta_k^c(k)| = \frac{A}{k} + \frac{2 B}{H_0}\left( 1 - \sqrt{a_d}\right).
\end{equation}
Thus, we can use this formula and calculate the collapse
time of the interesting values of $k$ we observe in the
cosmic microwave background (CMB), namely the range between
$10^{-3}$ Mpc$^{-1}$ $\leq k \leq 1$ Mpc$^{-1}$. These modes
covers the range of the multipoles $l$ of interest: $1 \leq
l \leq 2600$, were we made use of the relation\footnote{The
  relation between the angular scale $\theta$ and the
  multipole $l$ is $\theta \sim \pi/l$. The comoving angular
  distance, $d_A$, from us to an object of physical linear
  size $L$, is $d_A = L/(a\theta)$. $L/a \sim 1/k$, $d_A =
  R_D$ if the object is in the LSS, and using the first
  expression in this footnote, we get $l = k R_D $.} $\,l =
k R_D$. The collapse times for these modes can be regarded
as the times in which inhomogeneities and anisotropies first
emerged at the corresponding scales. These collapse times
are shown in figure (\ref{fig:tiempos_colapso}) for the best
values of $(A,B)$ given in the tables\footnote{ The reader
  should keep in mind that our parametrization of the
  inflationary regime has the conformal time running from
  large negative values to small negative values}
(\ref{tab:c_1}, \ref{tab:c_2}, \ref{tab:c_wigner}).

\begin{figure*}
  \centering \subfigure[$C_1$, with $A=10, B=1$.] {
    \label{fig:c1_e_foldings}
    \includegraphics[width=90mm,
    height=53mm]{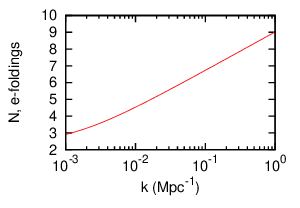}
  } \hspace{1cm} \subfigure[$C_2$, with $A=1000, B=1$.]{
    \label{fig:c2_e_foldings}
    \includegraphics[width=90mm,
    height=53mm]{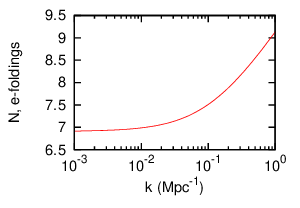}
  } \hspace{1cm} \subfigure[$C_{wigner}$, with $A=0.01,
  B=10$.]{
    \label{fig:cwigner_e_foldings}
    \includegraphics[width=90mm,
    height=53mm]{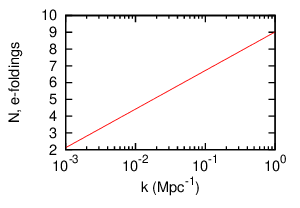}
  } \caption{Semi Logarithmic plot of the number of
    e-foldings between $a^H_k$ and $a^c_k$ for the three
    schemes, taking in account only the best values of
    $(A,B)$ in the range of $10^{-3}$ Mpc$^{-1}$ $< k < 1$
    Mpc$^{-1}$. For these plots $h = 0.7$.}
  \label{fig:e_foldings}
\end{figure*}

We can compare the value of the scale factor at the collapse
time $a(\eta_k^c)$, with the traditional scale factor at
``horizon crossing'' that marks the ``quantum to classical
transition'' in the standard explanation of inflation:
$a_k^H$. The ``horizon crossing'' occurs when the length
corresponding to the mode $k$ has the same size that the
``Hubble Radius'', $H_I^{-1}$, (in comoving modes $k=aH_I$)
therefore, $a_k^H \equiv a(\eta_k^H) = \frac{k}{H_I} =
\frac{3k}{8\pi G V}$. Thus the ratio of the value of scale
factor at horizon crossing for mode $k$ and its value at
collapse time for the same mode is
\begin{equation}
  \frac{a^H_k}{a^c_k} = {k\eta_k^c(k)} = {A + B R_D k} = A+Bl.
\end{equation}
Using the best-fit values for the different collapse
schemes, we can plot the e-folds elapsed between the modes
collapse and its horizon crossing. As we can see in the
figure (\ref{fig:e_foldings}) this quantity changes -at
most- of one order of magnitude in the range for $k$ for the
values of $A$ and $B$ that were considered more reasonable,
i.e. $a^H_k > a^c_k$, the time of collapse $\eta_k^c \simeq
10^{-3}\eta_k^H$ in this range.  The door is clearly open
for a more in detailed analysis and comparison to the the
actual empirical data, whereby one could hope to extract
robust information of the type discussed above.

\section{\label{sec:discussion}Discussion}
We have considered various, relatively \emph{ad hoc} recipes
for the form of the state of the quantum inflationary field,
that results, presumably from a gravitationally induced,
collapse of the wave function.  The breakdown of unitarity
that this entails, is thought to be associated with drastic
departures from standard quantum mechanics once the
fundamental quantum gravity phenomena come into play. We
have not discussed at any length this issue here and have
focused in the present treatment as purely phenomenological
aspects of the problem.

The analysis of the signatures of the different schemes of
collapse illustrate various generic points worth mentioning:
First, that, depending on the details of the collapse scheme
and its parameters, there can be substantial departures in
the resulting power spectrum, from the standard scale
invariant spectrum usually expected to be a generic
prediction from inflation. Of course it is known that there
exist other ways to generate modifications in the predicted
spectrum, such as considering departures from slow roll and
modifications of the inflaton potential and so forth.  In
the approach we have been following the modifications arise
from the details of a quantum collapse mechanism, a feature
tied to a dramatic departure from the standard unitary
evolution of quantum of physics that we have argued must be
invoked if we are to have a satisfactory understanding of
the emergence of structure from quantum fluctuations.  In
fact, by fitting the predicted and observational spectra,
these sort of modifications are possible sources of clues
about what exactly is the physics behind the quantum
mechanical collapse or whatever replaces it.  We saw that
generically one recovers the standard scale invariant
Harrison-Z\'eldovich spectrum if the collapse time (conformal
time) of the modes is such that $\eta^c_k k = constant$
\footnote{ This resembles the condition that is sometimes
  considered in the context of the so called trans-plankian
  problem.  There is however an important difference of what
  is supposed to occur at the (conformal) time that appears
  in this condition.  In addressing, the trans-plankian
  problem the time indicates when the mode actually comes
  into existence. In contrast , in our approach, the mode
  has existed always -modes are not created or destroyed-,
  but the state of the field in the corresponding mode {\bf
    changes (or jumps)} from the Adiabatic vacuum before the
  condition to the so called post-collapsed state after
  this, or a similar condition, is reached.}.
On the other hand and as shown in detail in \cite{Perez2006}
the simple generalization of the ideas of Penrose about the
conditions that would trigger the quantum gravity induced
collapse leads precisely to the such prediction for
$\eta^c_k $. We should however keep in mind that, even if
something of that sort is operating, the stochastic nature
of any sort of quantum mechanical collapse leads us to
expect that such pattern would not be followed with
arbitrarily high precision.  In this regard we have studied
the robustness of the various schemes in leading to an
almost scale invariant spectrum.  To this end we have
considered in this work, the simplest (linear) deviations
from the behavior of $\eta^c_k$ as a function of $k$ i.e.
we have explored in the three existing collapse schemes the
effects of having a time of collapse given by $\eta^c_k =
A/k +B R_D$.  The results of these studies are summarized in
figures \ref{fig:c1_log}, \ref{fig:c2_log},
\ref{fig:c_wigner_log} and tables \ref{tab:c_1},
\ref{tab:c_2} and \ref{tab:c_wigner}, so here we will only
point out one of the most salient features: We note that the
different collapse schemes lead to different types of
departures of the spectrum from the scale invariant one, for
instance the schemes $C_2(k)$ and $C_{wigner}(k)$ lead
naturally to a turning down of the spectrum as we increase
$l$.

It is worth noting that a turning down in the spectrum is
observed in the CMB data \cite{wmap2007},which is attributed
as a whole, in literature to \emph{the Damping
  Effect}\footnote{This effect basically is a damping for
  the photon density and velocity at scale $k$ at the time
  of decoupling by a factor of $e^{-k^2/k_D^2}$, where $k_D$
  is the diffusion scale and depends in the physics of the
  collisions between electrons and photons. Accordingly, the
  $C_l$ spectrum is also damped as $e^{-l^2/l_D^2}$ where
  $l_D \sim k_D d_A(\eta_d) \sim 1500$, for typical
  cosmological parameters.} , i.e. to the fact that
inhomogeneities are dampened do the non zero mean-free-path
of photons at that time of decoupling \cite{Damping}.  As
observed in the figures (\ref{fig:c2_log},
\ref{fig:c_wigner_log}) for some values of $(A,B)$ we obtain
an additional source of ``damping'' due to fluctuations in
the time of collapse about the pattern characterized by
$\eta^c_k k = constant$. It is expected that the PLANCK
probe will provide more information on the spectrum for
large values of $l$, so hopefully this characteristic of our
analysis could be analyzed and distinguished from the
standard damping in order to obtain interesting constraints
on the parameters $(A,B)$. In fact we believe that one
should be able to disentangle the two effects, because in
the cases in which our model leads to additional damping in
the spectrum, it also predicts that should be a rebound at
even higher values of $l$ (see figures \ref{fig:c1_log},
\ref{fig:c2_log}, \ref{fig:c_wigner_log}).

However, the most remarkable conclusion, illustrated by the
present analysis, is that by focussing on issues that could
be thought to be only philosophical and of principle, we
have been lead to the possibility of addressing issues
pertaining to some novel aspects of physics which could be
confronted with empirical observations. Further and more
detailed analysis based on direct comparisons with
observations are indeed possible, and should be carried out.
This together with the foreseeable improvements in the
empirical data on the spectrum, particularly in the large
$l$ region, and the large scale matter distribution studies,
should permit even more detailed analysis of the novel
aspects of physics that we believe are behind the origin of
structure in our universe.

\begin{acknowledgments}
  We would like to thank Dr. Jaume Garriaga for suggesting
  the consideration collapse scheme that follows the Wigner
  Function.  We acknowledge useful discussions on the
  subject with Dr. Alejandro Perez.  This work was supported
  by the grant DGAPA-UNAM IN119808 in part by a grant from
  DGEP-UNAM to one of the authors (AUT).
\end{acknowledgments}




\end{document}